\documentclass[pre,longbibliography,footinbib,twocolumn,preprintnumbers,amsmath,amssymb, superscriptaddress,floatfix]{revtex4-1}

\usepackage{amssymb}
\usepackage{amsmath}

\usepackage{color}
\usepackage[pdftex]{graphicx}

\usepackage{empheq}

\usepackage{mathtools}

\usepackage{makecell}

\usepackage[color=yellow]{todonotes}

\usepackage{enumerate}
\usepackage{mathtools}
\usepackage{stmaryrd}

\usepackage{enumitem}

\usepackage{textcomp}
\usepackage{gensymb}

\newcommand{\moy}[1]{\left\langle #1 \right\rangle}

\newcommand{\XX}[0]{\boldsymbol{X}}

\newcommand{\ex}[1]{\mathrm{e}^{#1}}

\newcommand{\dd}[0]{\mathrm{d}}
\newcommand{\ii}[0]{\mathrm{i}}

\newcommand{\FF}[0]{\boldsymbol{F}}
\newcommand{\ee}[0]{\boldsymbol{e}}
\newcommand{\rr}[0]{\boldsymbol{r}}

\newcommand{\RR}[0]{\boldsymbol{R}}
\newcommand{\xx}[0]{\boldsymbol{x}}
\newcommand{\qq}[0]{\boldsymbol{q}}

\newcommand{\ff}[0]{\boldsymbol{f}}

\newcommand{\pp}[0]{\boldsymbol{p}}

\newcommand{\kB}[0]{k_{\mathrm{B}}}

\newcommand{\nn}[0]{{\boldsymbol{n}}}

\newcommand{\cL}{\mathcal{L}}

\newcommand{\ind}[1]{_{\mathrm{#1}}}

\newcommand{\GG}{\boldsymbol{G}}

\newcommand{\xxi}{\boldsymbol{\xi}}

\newcommand{\gab}{g_{\alpha\beta}}
\newcommand{\kab}{k_{\alpha\beta}}

\newcommand{\uab}{u_{\alpha\beta}}

\newcommand{\Xab}{X_{\alpha\beta}}
\newcommand{\Yab}{Y_{\alpha\beta}}
\newcommand{\chiab}{\chi_{\alpha\beta}}
\newcommand{\tchiab}{\tilde\chi_{\alpha\beta}}
\newcommand{\phiab}{\phi_{\alpha\beta}}
\newcommand{\Phiab}{\phi_{\alpha\to\beta}}
\newcommand{\phiba}{\phi_{\beta\alpha}}
\newcommand{\Phiba}{\phi_{\beta\to\alpha}}

\definecolor{darkblue}{rgb}{0,0,0.6}
\definecolor{darkred}{rgb}{0.6,0,0}
\usepackage[colorlinks=true,urlcolor=darkblue,citecolor=darkblue,linkcolor=darkred]{hyperref}

\begin{document}

\title{Enhanced diffusion of tracer particles in nonreciprocal mixtures}

 \author{Anthony Benois}
 \affiliation{Sorbonne Universit\'e, CNRS, Physico-Chimie des \'Electrolytes et Nanosyst\`emes Interfaciaux (PHENIX), 4 Place Jussieu, 75005 Paris, France}

 \author{Marie Jardat}
 \affiliation{Sorbonne Universit\'e, CNRS, Physico-Chimie des \'Electrolytes et Nanosyst\`emes Interfaciaux (PHENIX), 4 Place Jussieu, 75005 Paris, France}

 \author{Vincent Dahirel}
 \affiliation{Sorbonne Universit\'e, CNRS, Physico-Chimie des \'Electrolytes et Nanosyst\`emes Interfaciaux (PHENIX), 4 Place Jussieu, 75005 Paris, France}

  \author{Vincent D\'emery}
 \affiliation{Gulliver, UMR CNRS 7083, ESPCI Paris PSL, 75005 Paris, France}
 \affiliation{Universit\'e Lyon, ENS de Lyon, Universit\'e Claude Bernard, CNRS, Laboratoire de Physique,
F-69342 Lyon, France}

 \author{Jaime Agudo-Canalejo}
 \affiliation{Department of Living Matter Physics, Max Planck Institute for Dynamics and Self-Organization, D-37077 G\"ottingen, Germany}

 \author{Ramin Golestanian}
 \affiliation{Department of Living Matter Physics, Max Planck Institute for Dynamics and Self-Organization, D-37077 G\"ottingen, Germany}
 \affiliation{Rudolf Peierls Centre for Theoretical Physics, University of Oxford, OX1 3PU, Oxford, UK}

 \author{Pierre Illien}
 \affiliation{Sorbonne Universit\'e, CNRS, Physico-Chimie des \'Electrolytes et Nanosyst\`emes Interfaciaux (PHENIX), 4 Place Jussieu, 75005 Paris, France}

\begin{abstract}
We study the diffusivity of a tagged particle in a binary mixture of Brownian particles with non-reciprocal interactions. Numerical simulations  reveal that, for a broad class of interaction potentials, non-reciprocity can  significantly increase the long-time diffusion coefficient of tracer particles, and that this diffusion enhancement is associated with a breakdown of the  Einstein relation. These observations are quantified and confirmed via two different and complementary analytical approaches: (i) a linearized stochastic density field theory, which is particularly accurate in the limit of soft interactions; (ii) a reduced two-body description, which is exact at leading order in the density of particles. The latter reveals that diffusion enhancement can be attributed to the formation of transiently propelled dimers of particles, whose cohesion and speed are controlled by the non-reciprocal interactions.

\end{abstract}

\date{\today}

\maketitle

\section{Introduction}

Intracellular functions are governed by the transport of ions, proteins, vesicles, or organelles, which are subject to strong thermal fluctuations, and which interact with each other through crowding, electrostatics and hydrodynamics.  In theoretical approaches, such systems are typically represented by a suspension of interacting particles, embedded in a solvent that causes their stochastic motion. These particles generally evolve very far from equilibrium, and are `active'
in the sense that they locally convert the chemical energy available in their environment into mechanical work.  
Even though a wealth of knowledge has been gathered on suspensions of single-species `polar' or self-propelled active particles \cite{Vicsek2012,Marchetti2013,Cates2015,Bechinger2016}, the reality is much more complex: suspensions of biological interest are generally strongly heterogeneous,
and made of particles without any established polarity on the considered timescales.

Very recently, `scalar' models for active matter, where agents are apolar but whose nonequilibrium dynamics results in spontaneous symmetry breaking, have been developed. For instance, one can consider   catalytic molecules, such as proteins or enzymes, that are involved in the production or consumption of smaller solute molecules. Each of them can  be seen as a local source or sink responding to the chemical gradients created by the other particles. When coarse-graining the degrees of freedom associated with solute molecules, the effective interactions between particles appear to break action-reaction symmetry \cite{Soto2014,Soto2015,Agudo-Canalejo2019,{Nasouri2020}}, and should be modeled as non-reciprocal. This line of research has recently gained a lot of importance, and now goes well beyond the interest for active colloids, with applications ranging from the design of new field theories \cite{Saha2020,You2020,Dinelli2022} and {advanced sampling techniques \cite{ghimenti2023sampling}}, to the interpretation of active matter experiments \cite{Yu2018,Grauer2020,Niu2018,{Meredith2020}}, and more generally phase transitions in nonequilibrium systems \cite{Fruchart2021}. Interestingly, mixtures of particles with non-reciprocal interactions can be mapped onto multi-temperature suspensions -- another class of scalar active matter that have received a lot of interest in the soft matter and biophysics communities \cite{Ganai2014,Grosberg2015,Tanaka2016a,Smrek2017,Ilker2020,Wang2020}. This mapping was formally established for Newtonian dynamics \cite{Ivlev2015}, and can be extended to stochastic overdamped dynamics (see Appendix \ref{app_mapping}).

The collective and structural properties of non-reciprocal mixtures have been studied rather extensively, revealing in particular their tendency to phase separate \cite{Agudo-Canalejo2019,Chiu2023,Mandal2022}. However, the properties of their fluctuations, as characterized by the dynamics of tracer particles (i.e.~individually-tracked, tagged particles) have been left aside so far, in spite of their importance. Indeed, the properties of tagged particles generally contain key information about the microstructure of the suspension and its small-scale dynamics \cite{{Hofling2013}}. They are also of  importance to quantify experiments that rely on single-particle tracking and allow accurate characterization of many intracellular processes \cite{Manzo2015}.

In this article, we study the diffusivity of a tagged particle in a binary mixture of particles with non-reciprocal interactions obeying overdamped Langevin dynamics. Brownian dynamics  simulations, together with two different analytical treatments of the stochastic dynamics,  reveal that non-reciprocity can  significantly increase the effective {long-time} diffusion coefficient of tracer particles. We measure the non-reciprocal contribution to its diffusivity:
\begin{equation}
D_\text{eff}= D_\text{recip} + \Delta D_\text{non-recip},
\end{equation}
 and, we show that, strikingly, this diffusion enhancement is associated with a breakdown of the  Einstein relation, which does not hold in this nonequilibrium case.  More precisely, the effective { long-time}  mobility can be written as $\mu_\text{eff} =D_\text{recip} /\kB T +  \Delta \mu_\text{non-recip} $, with the non-reciprocal correction being generally different from $\Delta D_\text{non-recip}/\kB T$.
We finally show that  diffusion enhancement can be attributed to the formation of transiently propelled dimers of particles, whose cohesion and speed are controlled by the non-reciprocal interactions.

\section{Model}

We consider a three-dimensional binary suspension of $N+1$ interacting particles, made of $N_A$ (resp. $N_B$) particles of species $A$ (resp. $B$), and one tracer particle (labeled $0$), that can either be of type $A$ or of type $B$. We denote by $\rho_\alpha=N_\alpha/V$ ($\alpha=A$ or $B$) the number density of each species (excluding the tracer), where $V$ is the volume of the system.  The overall density of bath particles is $\rho = N/V$, and $X_\alpha=\rho_\alpha/\rho$ is the fraction of $\alpha$ particles \footnote{{Note that taking the limit $X_A\to 0$ or $X_B\to 0$ corresponds to the situation of a single tracer coupled non-reciprocally to a bath of particles, which is covered by our approach.}}. We assume that each particle obeys an overdamped Langevin dynamics, in such a way that the evolution of the system is given by the set of coupled equations:
\begin{equation}
\label{overdampedLangevin}
\frac{\dd \rr_n}{\dd t} = \mu_{S(n)} \sum_{m\neq n} { \FF_{S(m)  \to S(n) }(\rr_n-\rr_m) }+ \sqrt{2 D_{S(n)}}\boldsymbol{\zeta}_{n}(t),
\end{equation}
 where $S(n)\in \{A,B\}$ denotes the species of particle $n$, and $\FF_{\beta  \to \alpha } (\rr)$ denotes the force exerted by a particle of species  $\beta$ on a particle of species $\alpha$  when the latter is located at $\rr$ relative to the former. 
 The bare diffusion coefficient of a particle of species $\alpha$ is related to the mobility $\mu_\alpha$ through the Einstein relation $D_\alpha = \kB T \mu_\alpha$, where $T$ is the temperature of the thermal bath in which the particles are embedded. 
 For simplicity, we will assume that all the particles have the same mobility $\mu_0$.
 The noise terms $\boldsymbol{\zeta}_n(t)$ have zero average and are delta-correlated: $\langle \zeta_{n,i}(t)\zeta_{m,j}(t') \rangle = \delta_{nm} \delta_{ij} \delta(t-t')$.
 
Importantly,  we assume that the interactions between  particles of different species can be non-reciprocal, namely that $\FF_{A\to B} (\rr)\neq -\FF_{B\to A}(-\rr)$. 
In order to probe the existence of enhanced diffusion in such a suspension, we  compute the {long-time}  diffusion coefficient of the tracer particle, defined as $D_\text{eff} = \lim_{t\to\infty} {\langle [ \rr_0(t)-\rr_0(0)]^2 \rangle}/{6t}$.

For simplicity, we write the forces as deriving from potentials (or `pseudo-potentials'): $ \FF_{\alpha\to \beta}(\rr_\beta-\rr_\alpha) = -\nabla_{\rr_\beta} \phi_{\alpha\to\beta} (|\rr_\alpha-\rr_\beta|)$. Note that we thus focus on divergence-free force fields. With this definition, the pseudo-potentials correspond to regular pair potentials when $\alpha=\beta$, but not otherwise.  The interactions between species can be defined through a matrix with elements $\Phi_{\alpha\beta} = \phi_{\alpha\to\beta}$ which is split between a symmetric (reciprocal) and antisymmetric (nonreciprocal) part:
\begin{align}
\boldsymbol{\Phi} &=\begin{pmatrix}
\phi_\text{rep}   &   \phi_\text{rep}   \\
\phi_\text{rep}+\phi_\text{att} &\phi_\text{rep}  \\
\end{pmatrix}\\
&= \begin{pmatrix}
  \phi_{A\to A}    & \phi_{AB}^\text{R}     \\
 \phi_{AB}^\text{R}    &  \phi_{B\to B}   \\
\end{pmatrix}
+\begin{pmatrix}
0   &   -\phi_{AB}^\text{NR}     \\
\phi_{AB}^\text{NR}  &0  \\
\end{pmatrix}.
\label{mat_Phi_split}
\end{align}
For concreteness, we assume that all the $(\alpha,\beta)$ pairs interact via a purely repulsive potential $\phi_\text{rep}(r)$, and that non-reciprocity is incorporated by assuming that the pseudo-potential $\phi_{A\to B}$ contains an additional attractive part $\phi_\text{att}(r)$ (in the notations of Eq.~\eqref{mat_Phi_split}, this means that $\phi_{AB}^\text{R} =  \phi_\text{rep} + {\phi_\text{att}}/{2} $ and  $\phi_{AB}^\text{NR} =  {\phi_\text{att}}/{2} $).

\section{Numerical evidence for enhanced diffusion}

{\subsection{Main results}}

We first present results from Brownian dynamics simulations, which consist in integrating the set of coupled overdamped Langevin equations [Eq.~\eqref{overdampedLangevin}], using a forward Euler-Maruyama scheme (see Appendix \ref{sec_app_methods}).  We consider different types of binary mixtures and the corresponding pair potentials, that  represent a broad range of physical situations (for each system, the expressions of $\phi_\text{rep}$ and $\phi_\text{att}$ are given in Table \ref{table_potentials}): (i) suspensions of hard particles with short-range repulsion given by the Weeks-Chandler-Andersen potential and long-range Lennard-Jones attraction \cite{Chiu2023,Mandal2022}; (ii) particles with softcore interactions, modeled by a `Gaussian' potential, which are relevant to describe the interactions between polymer coils \cite{Louis2000,{Likos2001,Lang2000,Louis2000}}; and (iii) Yukawa interactions, that represent screened Coulomb interactions or may arise from `chemical interactions' between diffusiophoretic colloids \cite{Soto2014,Soto2015,Agudo-Canalejo2019,Meredith2020} (the ranges $\lambda$ and $\lambda'$ of the Yukawa-like $\phi_\text{rep}$ and $\phi_\text{att}$ are chosen such that $\phi_{A \to B}$ has an attractive part). For all these systems, the energy parameters $\varepsilon$ and $\delta$ represent respectively the strength of the repulsive and attractive parts of the potentials. When non-reciprocity is very strong, the suspension may be unstable and phase separate -- this effect was for instance evidenced in suspensions of colloids with chemically-mediated \cite{Agudo-Canalejo2019} or LJ-WCA \cite{Chiu2023,Mandal2022} interactions.
However, we emphasize that all our simulations are performed in the homogeneous regime, where the non-reciprocal mixture does not display any phase separation.

\begin{figure*}
\begin{center}
\includegraphics[width=1.99\columnwidth]{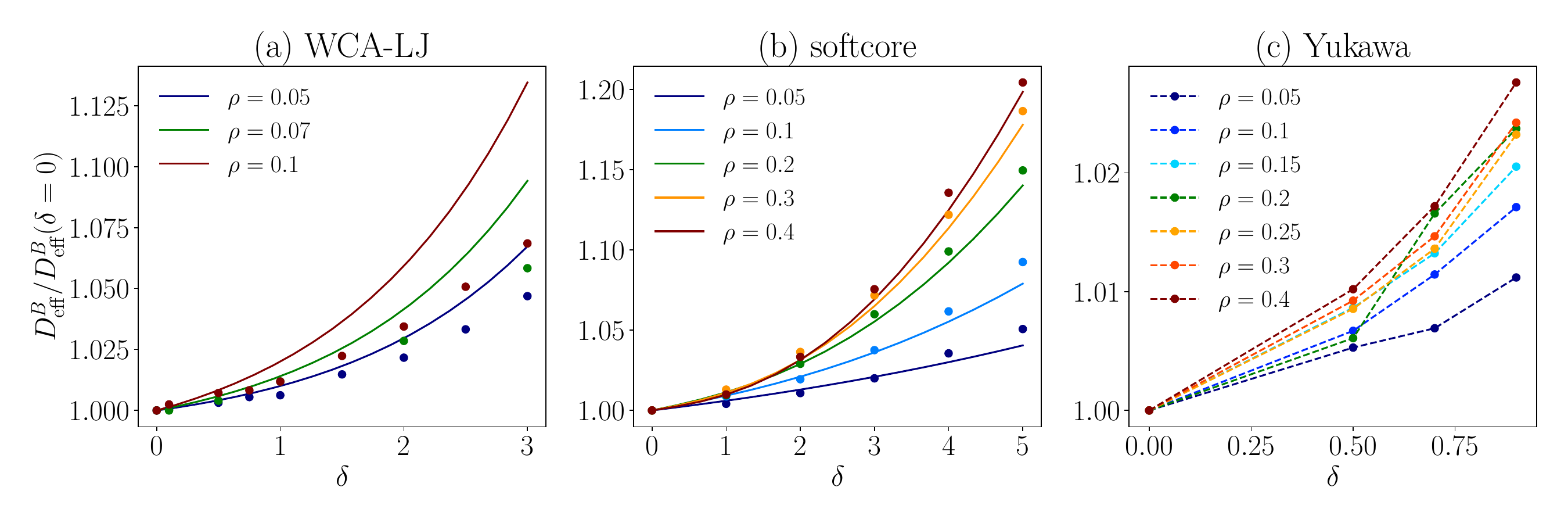}
\caption{Long-time  diffusion coefficients of  $B$ particles   (rescaled by their value in the reciprocal case, $\delta=0$) as a function of the parameter $\delta$, which quantifies the intensity of non-reciprocity. Throughout the paper, energies are  measured in units of $\kB T$ and distances in units of $\sigma$, the diameter of the particles. In all  simulations, and unless otherwise specified, $\rho_A=\rho_B=\rho/2$. {(a) solid lines are  analytical predictions in the low-density limit (Appendix \ref{app_low_dens});} (b) solid lines are  analytical predictions in the limit of soft interactions [Eq.~\eqref{main_result_Deff}]; (c) dashed lines are guides to the eye. See Table \ref{table_potentials} for the expressions of the pair potentials (parameters: $\varepsilon=1$, $\sigma=1$, $\lambda=1$, $\lambda'=1.7$). }
\label{fig_Deff_B}
\end{center}
\end{figure*}

In each of these systems, we measure numerically the diffusion coefficient of tagged $B$ particles, as summarized in Fig.~\ref{fig_Deff_B}.   For all three sets of simulations, diffusion is enhanced as non-reciprocity increases (a similar effect is observed for tagged $A$ particles, see Appendix~\ref{app_additional_results}). In the particular case of the softcore potentials, the relative enhancement can reach values as high as 20\% \footnote{\textcolor{blue}{Note that the enhancement due to non-reciprocity can actually compensate the hindering due to crowding, in such a way that $D_\text{eff}^B$ may exceed its bare value (i.e. its value in the limit of infinite dilution) denoted by $D_B$.}}. The radial distribution functions reveal a strong pairing between $A$ and $B$ particles (Appendix~\ref{app_additional_results}), which is interpreted as a consequence of the `predator-prey' dynamics that emerges from non-reciprocal interactions: $B$ particles chase $A$ particles while $A$ tend to run away from $B$, resulting in enhanced dynamics at the scale of tagged particles -- this effect will be described in Section \ref{sec_low_dens}.

\begin{table}
  \centering 
  \begin{tabular}{c|c|c}
   & $\phi_\text{rep}(r)$ &  $\phi_\text{att}(r)$\\   \hline
  LJ-WCA &  $ 4\varepsilon \left[ \left( \frac{\sigma}{r}\right)^{12}-\left( \frac{\sigma}{r}\right)^{6} \right]    \theta(2^{1/6}\sigma-r) $ & $ 4\delta \left[ \left( \frac{\sigma}{r}\right)^{12}-\left( \frac{\sigma}{r}\right)^{6} \right]   $  \\
softcore   & $\varepsilon \ex{-(r/\sigma)^2}$ & $-\delta \ex{-(r/\sigma)^2}$ \\
Yukawa & $\varepsilon \frac{\sigma}{r}\ex{-r/\lambda}$ & $-\delta \frac{\sigma}{r}\ex{-r/\lambda'}$ 
\end{tabular}
  \caption{Expressions of the repulsive and attractive part of the interaction potentials considered in the  simulations.}
  \label{table_potentials}
\end{table}

{\subsection{Additional comments}}

{We emphasize that choosing another decomposition of the matrix $\mathbf{\Phi}$ [Eq. \eqref{mat_Phi_split}] is not expected to affect the main results. Non-reciprocity is controlled by the intensity of the pseudo-potential $\phi_\text{att}$ (through the parameter $\delta$), which is independent of the decomposition into a reciprocal and a non-reciprocal part. There are possibly other choices of the decomposition that would surely modify the `enhancement', if we change the reference with respect to which diffusion coefficient is measured, but we argue that, in the present situation, the best choice to rationalize our simulation results is to take $D_\text{eff}(\delta=0)$ as a reference. }

{Finally, it is interesting to think about the relevant observables that should be used to discriminate between reciprocal or non-reciprocal interactions in experimental measurements. As the diffusion coefficient of a tagged particle can be larger or smaller depending on the sign of $\delta$, we believe that the diffusion coefficient would probably not be a sufficient observable to identify the reciprocal case. The measurement of diffusion coefficients could be completed by observables that characterize the structure and spatial organization of the system, in which the signature of the pairing mechanism can be observed (see for instance the radial distribution functions shown on Fig. \ref{fig_gr}).}

\section{Breakdown of the Einstein relation}

\begin{figure}
\begin{center}
\includegraphics[width=\columnwidth]{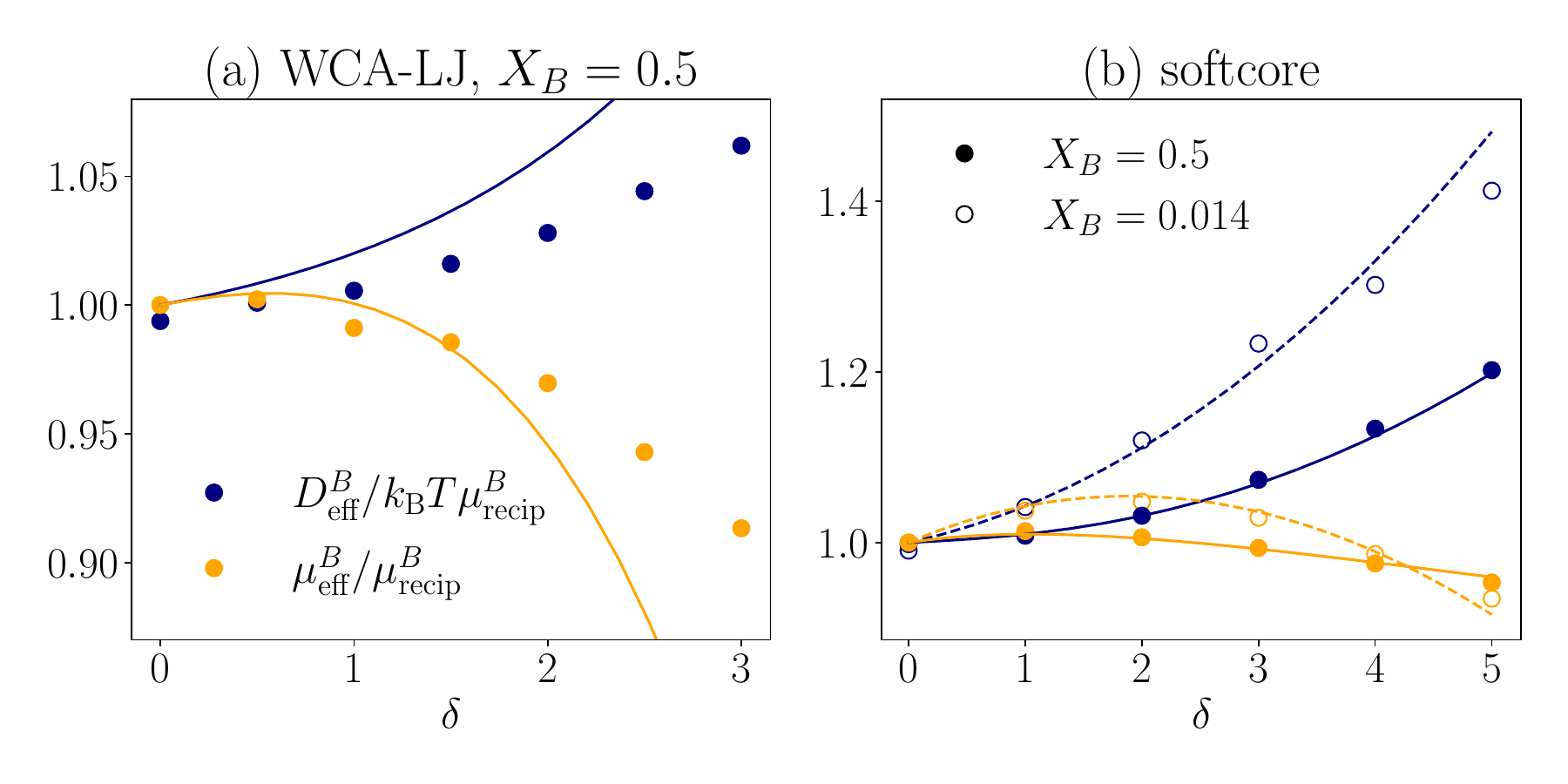}
\caption{{ Long-time  diffusion coefficients and mobilities of tagged $B$ particles  in non-reciprocal mixtures as a function of the parameter $\delta$. Densities are (a) $\rho=0.10$; (b) $\rho=0.40$ (see caption of Fig.~\ref{fig_Deff_B} for the other parameters). In both panels, symbols are results from Brownian dynamics simulations. (a) Solid lines are analytical predictions in the low-density limit (Appendix \ref{app_low_dens}). (b) Solid and dashed lines are analytical predictions in the limit of soft interactions [see Eq.~\eqref{main_result_Deff} for $D_\text{eff}^B$ and Eq. \eqref{mueff_soft_general} for $\mu_\text{eff}^B$]. }}
\label{Einstein_breakdown}
\end{center}
\end{figure}

In order to probe the validity of the Einstein relation in such mixtures, we measured the mobility of tagged particles, aiming at comparing it to the effective diffusion coefficient defined earlier. To this end, in the numerical simulations, we add a constant external force $\ff = f\ee_x$ to the tagged particle, and measure its mobility defined as $\mu_\text{eff} = \lim_{f \to 0} \langle v_x \rangle /f$, where $\langle v_x \rangle$ is the average velocity attained by the tagged particle along direction $x$, in the stationary limit. For an equilibrium system, the effective mobility of the tracer is expected to be related to its effective diffusion coefficient through the Einstein relation $D_\text{eff} = \kB T \mu_\text{eff} $. We compare in Fig.~\ref{Einstein_breakdown} the effective diffusion coefficients and mobilities as measured from simulations: the increasing mismatch between their values as $\delta$ increases is a clear indication of the breakdown of the Einstein relation in this nonequilibrium situation.

\section{Analytical description in the limit of soft interactions}

In order to quantify these phenomena and to offer analytical insight, we coarse-grain the dynamics and define the density of bath particles of species $\alpha$ as $\hat\rho_\alpha(\rr,t) = \sum_{n\neq0,S(n)=\alpha}\delta (\rr_n(t)-\rr) $, where the sum runs over all the particles of species $\alpha$ except the tracer (if of species $\alpha$), so that the tracer is `taken out' of the definition of the densities \cite{Demery2014}. Using It\^o calculus \cite{Gardiner1985}, and relying on the usual derivation proposed by Dean for a single-component fluid \cite{Dean1996} and later extended to binary mixtures \cite{Poncet2016}, we obtain the coupled equations for the fields $\hat \rho_\alpha$:
\begin{align}
\label{Dean_A}
&\partial_t \hat\rho_\alpha    =\sqrt{2D_0} \nabla \cdot [\boldsymbol{\eta}_\alpha \sqrt{\hat\rho_\alpha}] + D_0 \nabla^2\hat \rho_\alpha    \\
&-\mu_0 \nabla \cdot \Big[  \hat \rho_\alpha  \sum_{\beta \in \{A,B\}} {\FF_{\beta\to\alpha}}\ast \hat\rho_\beta +\hat\rho_\alpha  {\FF_{ S(0)\to \alpha}}\ast \delta_{\rr_0} \Big]  \nonumber
\end{align}
with the space-dependent noise $ \eta_{\alpha,i} (\rr,t)$ of average zero, and with correlations $
\langle \eta_{\alpha,i} (\rr,t)\eta_{\beta,j}(\rr',t')\rangle = \delta_{\alpha\beta} \delta_{ij} \delta(\rr-\rr') \delta(t-t')$.
In Eq.~\eqref{Dean_A}, the symbol $\ast$ represents spatial convolution: $(f\ast g)(\rr) = \int \dd \rr' \; f(\rr') g(\rr-\rr')$, and we use the shorthand notation $\delta_{\rr_0}(\rr)=\delta(\rr-\rr_0)$.   The evolution of the tracer position is given by the overdamped Langevin equation, Eq.~\eqref{overdampedLangevin}, written for $n=0$. 

Although explicit, this joint description of the tracer-bath dynamics is quite complicated, as it involves nonlinear couplings and multiplicative noise. The dynamics of the fields can be solved perturbatively by linearizing around the homogeneous state of density $\rho_\alpha$, and assuming $|\hat\rho_\alpha - \rho_\alpha| \ll \rho_\alpha$ \cite{Demery2014,{Poncet2016}} . 
To treat the nonlinear coupling between the fields and the position of the tracer $\rr_0(t)$,  we rely on a path-integral formulation \cite{Demery2011} that we recently extended to the case of a binary mixture \cite{Jardat2022}. We finally reach an  expression for the {long-time}  diffusion coefficient of the tracer particle as a Fourier integral:
\begin{align}
&\frac{ {D}_\text{eff} }{D_0} = 1-\sum_{\substack{\alpha,\beta,\gamma \\ \in \{A,B\}}}\sqrt{X_\alpha X_\gamma} \int_0^\infty \frac{\dd q}{6\pi^2} \rho q^2  \tilde \phi _{\alpha\to S(0)}(q) \nonumber \\
& \times \left[ C_{S(0)\to\gamma}^{\alpha\beta\gamma} \tilde \phi _{S(0)\to\gamma}(q)  +C_{\gamma \to S(0)}^{\alpha\beta\gamma}     \tilde \phi _{\gamma \to S(0)}(q)   \right]
\label{main_result_Deff}
\end{align}
where the tildes represent Fourier transforms, and where the functions $C_{S(0)\to\gamma}^{\alpha\beta\gamma}(q)$ and $C_{\gamma \to S(0)}^{\alpha\beta\gamma}(q)$ are given in Appendix \ref{app_lin_Dean} in terms of the densities of each species, their interaction potentials, and their mobilities {(note that, in that Appendix, we actually consider the more general situation where the tracer can be a different species than A or B). 
Additionally, the long-time effective mobility of the tracer particle can be computed with similar tools:
technically, this is done by applying a small external force $\ff$ to the tracer and computing the correction to the average tracer velocity (Appendix \ref{app_corr_mob}) \cite{Demery2014}.
}

Eq.~\eqref{main_result_Deff} is one of the main analytical results of this article, and several comments follow: (i) Up to a numerical integration, the effective {long-time}  diffusion coefficient is obtained as an explicit expression in terms of all the parameters of the problem; (ii) In our formalism, one can actually find a more general expression of the effective {long-time}  diffusion coefficient, for cases in which $A$ and $B$ particles have different mobilities and are connected to different thermostats (Appendix \ref{app_lin_Dean}); and (iii) This result should be understood as a perturbative expansion in the limit of weak interactions between the particles. Therefore, when compared to our numerical results for the different interaction potentials considered in Table \ref{table_potentials}, it is only valid for the softcore interactions:
the agreement between our analytical theory and numerical simulations is very good [Fig.~\ref{fig_Deff_B}(b)]. Strikingly, it shows that this linearization procedure remains true even very far from equilibrium.

 In order to discuss some consequences of Eq.~\eqref{main_result_Deff}, we consider a simpler situation, where the tagged particle is coupled in a non-reciprocal way to a single bath (we will assume that the probe is of species $B$ and the bath particles of species $A$). We take $\phi_{A\to A}(r)=\phi_{B\to A}(r)=v(r)$ and $\phi_{A\to B}(r)=(1-\delta) v(r)$, in such a way that $\delta$ measures non-reciprocity, just like in our numerical simulations. In this case, the effective { long-time}  diffusion coefficient of the tagged particle has the simple expression:
 \begin{equation}
 D_\text{eff}= D_\text{recip}+D_0 \int \frac{\dd \qq}{(2\pi)^3} \frac{\rho\delta(2+\delta\rho\tilde{v})\tilde{v}^2}{3(1+\rho\tilde{v})(2+\rho \tilde v)^2},
\end{equation}
where $\frac{D_\text{recip}}{D_0}=1-\frac{\rho}{3} \int \frac{\dd\qq}{(2\pi)^3} \frac{\tilde{v}^2}{(1+\rho\tilde{v})(2+\rho \tilde v)}$. The effective mobility of the tracer can be computed by assuming that it is driven by a harmonic trap with vanishing stiffness, and by adapting earlier calculations \cite{Demery2019}. We find that the effective mobility is given by 
\begin{equation}
\mu_\text{eff} = \frac{D_\text{recip}}{\kB T} +\mu_0 \int \frac{\dd \qq}{(2\pi)^3}  \frac{\rho\delta(3-\delta+\delta\rho\tilde{v})\tilde{v}^2}{3(1+\rho\tilde{v})(2+\rho \tilde v)^2},
\label{mueff_oneB}
\end{equation}
which we compare to numerical data in Fig.~\ref{Einstein_breakdown}(b). 
Therefore, this confirms the breakdown of the Einstein relation, which is only retrieved in the reciprocal case $\delta=0$ and the trivial case $\delta=1$, where the bath has no effect on the tracer.

\section{Low-density limit}
\label{sec_low_dens}

We finally consider the low-density limit of the problem, where it  actually reduces to a two-particle situation: one of them is the tagged particle, the other one is a  bath particle. To ease the notation, we will assume that the tracer, at position $\rr_0$, is of species $\alpha$, and the considered bath particle, at position $\rr_\text{b}$, is of species $\beta$.
The pair correlation of the tracer with the bath particle is obtained by solving the Smoluchovski equation for the two-body probability density $P_{\alpha\beta}(\rr_0,\rr_\text{b},t)$.
Using the variables $\rr = \rr_\text{b}-\rr_0$ and $\RR=(\rr_0+\rr_\text{b})/2$, it reads
\begin{align}\label{eq:low_dens_r_R}
&\partial_tP_{\alpha\beta} = 2D_0\nabla_{\rr}^2 P_{\alpha\beta}+\mu_0\nabla_{\rr}\cdot \left[P_{\alpha\beta}\nabla_{\rr}(\Phiab +\Phiba) \right] \nonumber\\
&+\frac{D_0}{2}\nabla_{\RR}^2 P_{\alpha\beta}+\frac{\mu_0}{2}\nabla_{\RR} P_{\alpha\beta}\cdot\nabla_{\rr}(\Phiba-\Phiab).
\end{align}
Integrating over the position of the center of mass $\RR$, and defining $\uab(r)= [\Phiab(r)+\Phiba(r)]/2\kB T$,  we find the stationary solution of Eq.~\eqref{eq:low_dens_r_R}: $g_{\alpha\beta}^0(r) = \exp[-\uab(r)]$. Interestingly, this is analogous to the simple equilibrium pair distribution in the low-density limit, but with the interaction potential taken as the average between the two non-reciprocal pseudo-potentials. The {long-time}  diffusion coefficient of the tracer, as well as its effective mobility, can be computed using standard methods  \cite{Lekkerkerker1984,Ilker2021}, and the comparison between these two observables shows that the Einstein relation does not hold in this limit either (see Appendix \ref{app_low_dens}). {Although the potentials used in this study do not allow the derivation of explicit expressions for the long-time diffusion coefficient and mobility in this limit, we can evaluate through the numerical resolution of ordinary differential equations. For LJ/WCA potentials, the results are shown on Fig. \ref{fig_Deff_B}(a), Fig. \ref{Einstein_breakdown}(a) and Fig. \ref{fig_Deff_A}(a).}

This low-density approach also reveals that, if the effective potential $\uab(r)$ has a deep enough minimum, an $\alpha\beta$ pair may form a `transient dimer' that remains bound for some time. Indeed, the dynamics of their center of mass $\RR$ can be read from the effective equation of motion $\dot\RR(t) = \frac{\mu_0}{2}\nabla(\Phiab-\Phiba)+\sqrt{D_0}\xxi(t)$, where $\xxi(t)$ is a Gaussian white noise with unit variance.
If the interaction is non-reciprocal, the first term on the r.h.s.~is non-zero and represents a self-propulsion term, which depends solely on the inner variable $\rr(t)$.
The characteristics of self-propulsion depend on the shape of the potentials: (i) If the minimum of $\uab(r)$ is at $r=0$ and the potentials behave as $\phiab(r)\sim \kab r^2/2$ around $r=0$, then the dynamics of $\rr(t)$ is linear,  $\rr(t)$ is an Ornstein-Uhlenbeck process, and the coordinate $\RR(t)$ therefore behaves as an  Active Ornstein-Uhlenbeck particle \cite{Szamel2014}; 
(ii) On the contrary, if the minimum of $\uab(r)$ is at $r^*>0$, then the modulus of  interparticle vector remains confined close to $r^*$, and one can define a  self-propulsion velocity $V_0 \simeq \frac{D_0}{2} [\phiab'(r^*)-\phiba'(r^*) ] $: the coordinate $\RR(t)$  behaves as an Active Brownian Particle \cite{Romanczuk2012}, with a rotational diffusion coefficient $D_r \simeq 2D_0/{r^*}^2$.

In our numerical simulations, when the overall density $\rho$ is small enough, we observe that $D^A_\mathrm{eff}\simeq D^B_\mathrm{eff}$, even for strong non-reciprocity (Appendix \ref{app_additional_results}). In contrast, at higher densities, these two values differ more clearly.   This supports the idea that, at low density, diffusion enchancement can be related to the pairing between $A$ and $B$ particles. This effect is reminiscent of the self-propelled dimers observed in very dilute suspensions of chemotactic colloids~\cite{Soto2014, Yu2018,Niu2018,Meredith2020}. {This relationship between the nonreciprocal mixture and suspensions of active particles (Brownian or Ornstein-Uhlenbeck) could be used to define an `active temperature' and a generalized Einstein relation \cite{Solon2022}}.

\section{Discussion}

In this article, we showed that non-reciprocal interactions between Brownian particles could significantly enhance their diffusivity. Non-reciprocity, which plays a predominant role in the interaction between chemically active particles, is thus expected to have a significant impact on the efficiency of molecular transport, and on the kinetics of diffusion-limited reactions.  These observations, together with the mapping between non-reciprocal and two-temperature mixtures,  open the way  to the interpretation of the rich phenomenology of non-reciprocal and multi-temperature mixtures, and to the local structures that emerge from the local energy transfers at the microscopic scale \cite{Schwarcz2023}. In the biological context, we believe that these concepts could find their applications in elucidating the role played by ATP-fueled activity in the fluidization of the intracellular medium, and of its rheological properties~\cite{Parry2014}.

\section*{Acknowledgments}

The authors acknowledge Roxanne Berthin for her assistance with the computational resources of PHENIX laboratory, and Rodrigo Soto for discussions.

{
\section*{Data availability}

The input files and raw data used for the figures are available on Zenodo (\url{https://doi.org/10.5281/zenodo.10126206}).

}

\appendix

\section{Mapping from multi-temperature suspensions to non-reciprocal mixtures}
\label{app_mapping}

The mapping from multi-temperature suspensions to non-reciprocal mixtures was demonstrated for a binary, underdamped suspension in Ref.~\cite{Ivlev2015}. We extend here these arguments for an arbitrary number of species, in the overdamped limit. We write the generic  Smoluchowski equation for a suspension of $N$ particles connected to $N$ different thermostats: 
\begin{align}
&\partial_t \mathcal{P}(\rr^N;t) = \sum_{n=1}^N \left \{ \kB T_n \mu_n \nabla_{\rr_n}^2\mathcal{P}(\rr^N;t) \right.\nonumber\\
&\left.- \mu_n \nabla_{\rr_n}\left[\mathcal{P}(\rr^N;t)\sum_{m\neq n} \FF (\rr_n-\rr_m) \right] \right\},
\label{Smolu}
\end{align}
where $\mu_n$ is the mobility of particle $n$, $T_n$ its temperature, $\FF$ is a reciprocal force field (reciprocal in the sense that it depends only on the distance between the particles, and not on their species), and $\mathcal{P}(\rr^N;t)$ is the $N$-body probability distribution (we use the shorthand notation: $\rr^N=(\rr_1,\dots,\rr_N)$). It is straightforward to show that this equation also describes the dynamics of a suspension of $N$ particles connected to a single thermostat $T$, but with mobilities $\mu'_n=(T_n/T)\mu_n$, and interacting via `non-reciprocal forces' $\FF_{m\to n} (\rr_n-\rr_m) = (T/T_n) \FF (\rr_n-\rr_m)$ (here $\FF_{m\to n}$ denotes the force exterted by particle $m$ on particle $n$). Therefore, a mixture of overdamped particles with multiple temperatures can be mapped onto a mixture with non-reciprocal forces, but we underline that the converse is not always true. This gives additional justification for the study of non-reciprocal mixtures, that appears to be more general, and that they can give insight into the physics of multi-temperature systems.

\section{Numerical methods}
\label{sec_app_methods}
To perform Brownian dynamics simulations we have used the LAMMPS computational package \cite{lammps_web,PLIMPTON19951,LAMMPS}.   We used the  command `fix brownian' that allows one to integrate overdamped Langevin equations for the positions of particles thanks to a forward Euler-Maruyama scheme. To allow the forces between $A$ and $B$ particles to be non-reciprocal, we have used Pylammps, the wrapper python class for LAMMPS.  More precisely, we have added in the simulation box a number of `ghost' particles (named $C$ in what follows) equal to that of $A$ particles. The interaction potential between $A$ and $B$ is the interaction potential $\phi_{\rm rep}(r)$, and  $C$ and $B$ particles undergo the attractive interaction potential $\phi_{\rm att}(r)$ (see Table \ref{table_potentials}). 
At each time step of the simulation, ghost particles $C$ are put at the exact same positions as particles $A$, so that they exert an additional attractive force  on particles $B$ as if they were $A$ particles. This additional force does not influence particles $A$: as a result, the total force between $A$ and $B$ particles are non-reciprocal. 
 To compute interaction forces, cutoff distances equal to $2.5\sigma$, $2.5\sigma$,  and $5\sigma$ are used for the Lennard-Jones, softcore and Yukawa interaction potentials, respectively.  The input mobility of particles is the same for $A$ and $B$ particles. 
 
In every case, a total number of $N_A=2000, N_B=2000, N_C=2000$ particles are placed in a cubic simulation box with periodic boundary conditions. The length of the box $L_\text{box}$ is varied to change the total density $\rho=(N_A+N_B)/L_\text{box}^3$ of the system. The time step varies between $\Delta t=0.0002t^\star$ and $\Delta t=0.002t^\star$, depending on the interaction potential  and of the density of the system, with $t^\star=\sigma^2/(k_{\rm B}T\mu_0)=\sigma^2/D_0$ the time needed for a particle to diffuse over a length equal to its size. In each case, we begin by one long trajectory of $10^7$ time steps to equilibrate the system at $\delta=0$ (reciprocal case). For each value of $\delta$, starting from a configuration representative of the equilibrium situation ($\delta=0$), one long trajectory of about  $10^7$ time steps is run to reach a stationary state, characterized by constant radial distribution functions. Then, mean squared displacements of tracers are averaged over three to nine independent trajectories of about $10^7$ time steps (depending on the time step and density), and over particles and time. The mean-squared displacements were found to be linear at all times for every system investigated here. The uncertainty of the computed self-diffusion coefficients was evaluated from the standard deviation of values obtained from different trajectories. The uncertainty on ${D}_\text{eff}/{D_0}$ was in each case smaller than $0.005$. Note that the size of the symbols used in the figures is larger than these error bars.

Finally, to compute the mobility, we have added a force on $50$ tracer $B$ particles chosen at random. The amplitude of the force is chosen to ensure that we stay in the linear regime, {i.e.} the displacement with time is proportional to the force. Starting from equilibrated configurations obtained at each $\delta$, we have run simulations of about $2\times 10^6$ time steps with the added force. The mobility is computed from $\langle[\mathbf{r}(t) - \mathbf{r}(t=0)]/{t}\rangle$, averaging over the last $20\%$ steps of three independent trajectories, and over the $50$ tracer particles.

\section{Additional results from Brownian dynamics simulations}
\label{app_additional_results}

In this Section, we present additional numerical results.

\begin{figure}
\begin{center}
\includegraphics[width=\columnwidth]{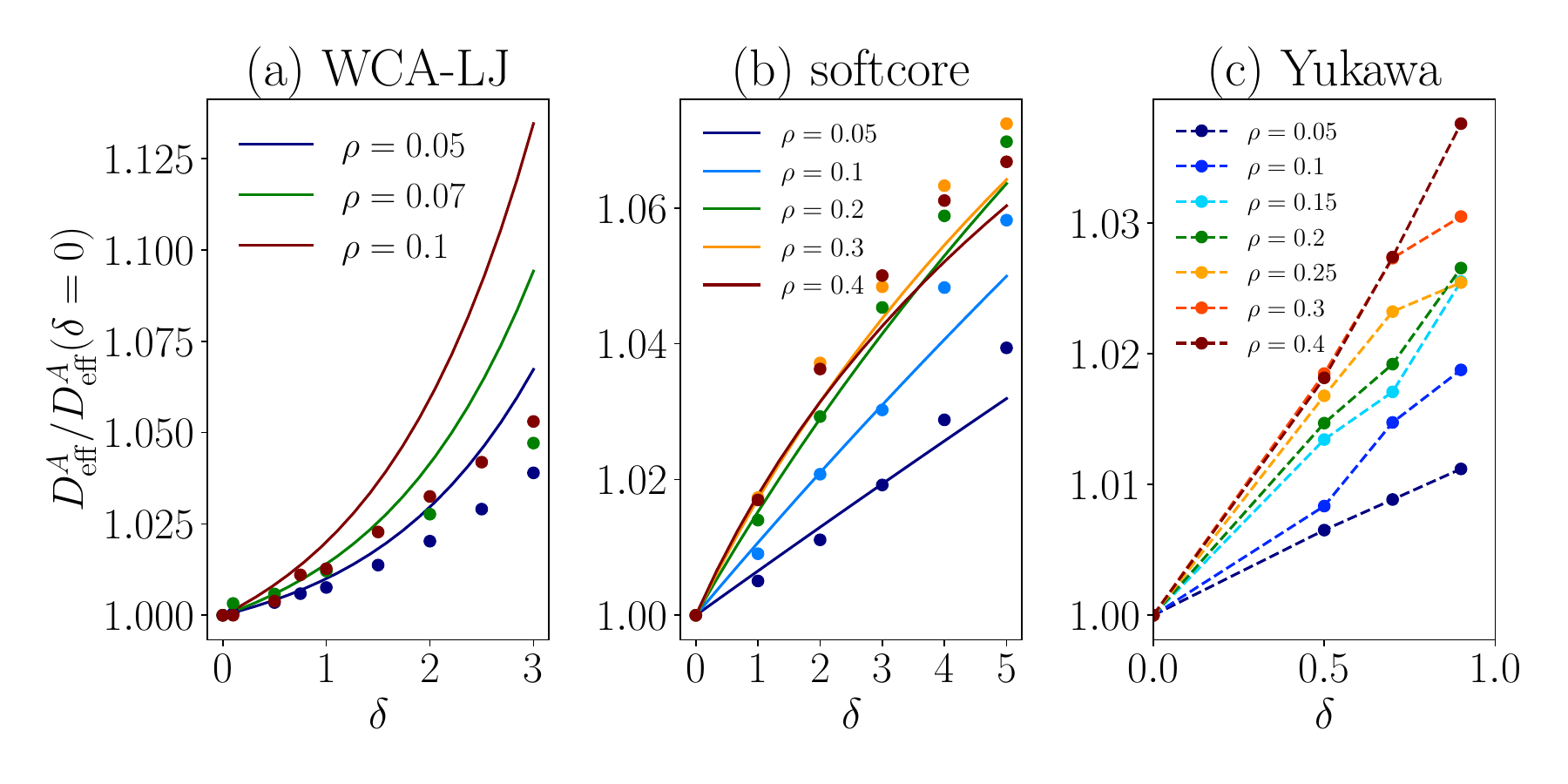}
\caption{Long-time diffusion coefficients of tagged $A$ particles  in non-reciprocal mixtures (rescaled by their value in the reciprocal case, $\delta=0$) as a function of the parameter $\delta$, which quantifies the intensity of non-reciprocity. Throughout the paper, energies are  measured in units of $\kB T$ and distances in units of $\sigma$, the diameter of the particles. In all  simulations, $\rho_A=\rho_B=\rho/2$.  {(a) solid lines are  analytical predictions in the low-density limit (Appendix \ref{app_numerical_ode});} (b) solid lines are  analytical predictions in the limit of soft interactions [Eq.~\eqref{main_result_Deff}]; (c) dashed lines are guides to the eye. See {Table \ref{table_potentials}} for the expressions of the pair potentials (parameters: $\varepsilon=1$, $\sigma=1$, $\lambda=1$, $\lambda'=1.7$).}
\label{fig_Deff_A}
\end{center}
\end{figure}

\begin{figure}
\begin{center}
\includegraphics[width=\columnwidth]{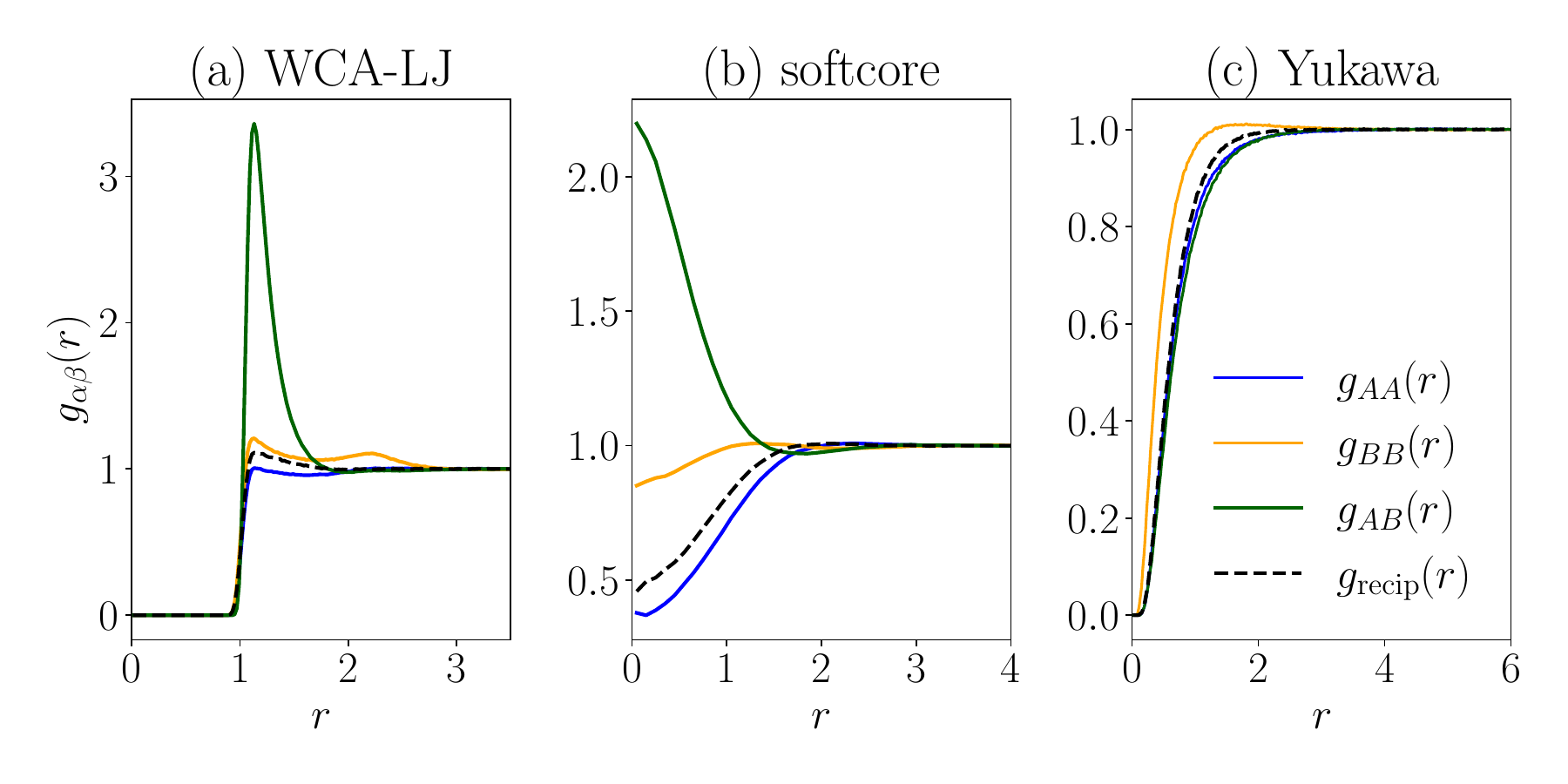}
\caption{Radial distribution functions for the different pairs of species in the binary mixtures ($AA$, $BB$, $AB$), and for the different interaction potentials used in the Brownian dynamics simulations (WCA-LJ: $\rho=0.1$, $\delta=3$, softcore: $\rho=0.3$, $\delta=5$, Yukawa: $\rho=0.3$, $\delta=0.9$). The curves for $g_\text{recip}(r)$ correspond to the $AA$ radial distribution functions in the reciprocal case (the $BB$ and $AB$ radial distribution functions are identical within statistical noise, as expected in this situation).}
\label{fig_gr}
\end{center}
\end{figure}

\begin{figure}
\begin{center}
\includegraphics[width=\columnwidth]{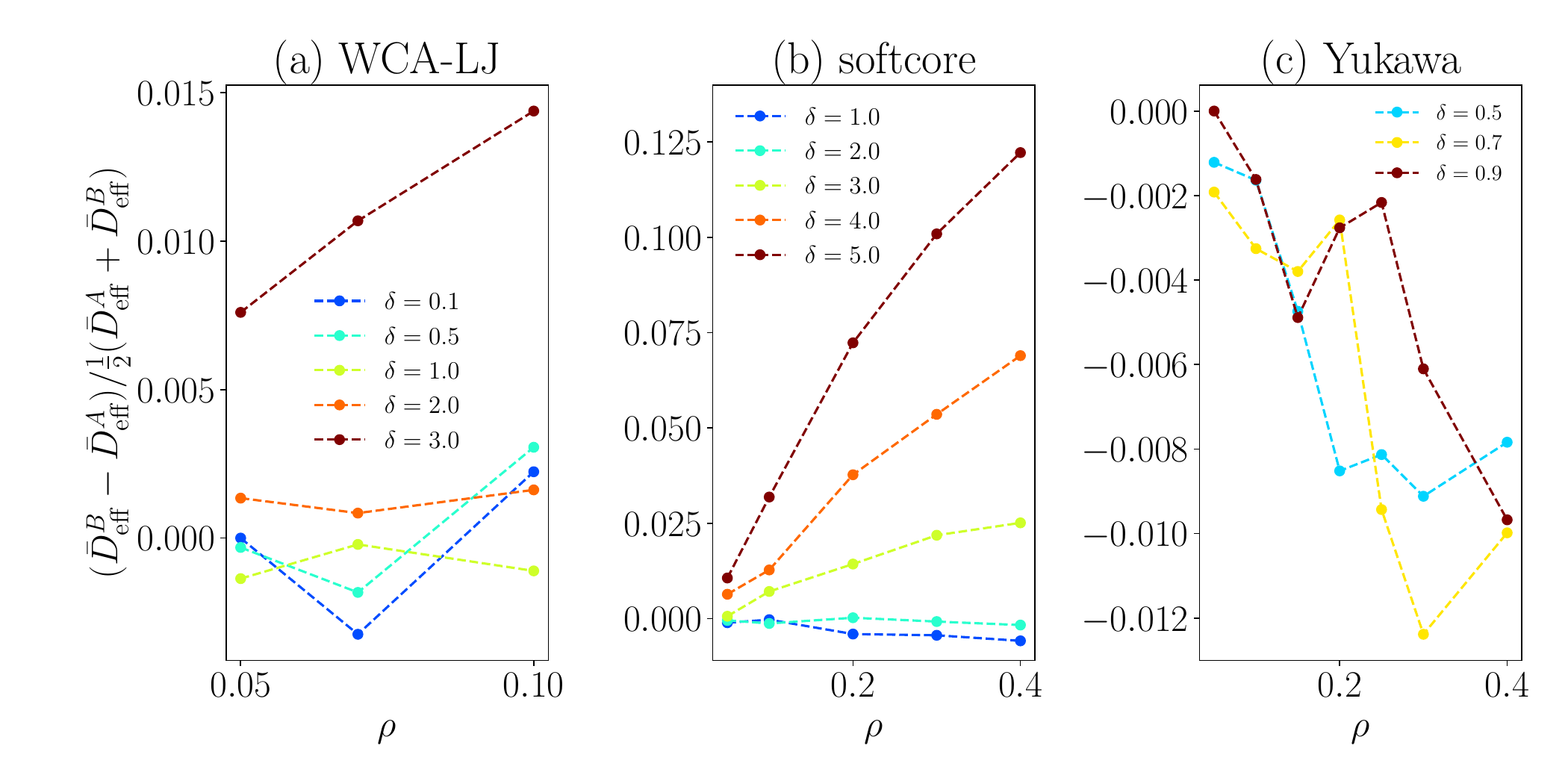}
\caption{Relative difference between the effective diffusion coefficient of $A$ and $B$ particles, for the different set of numerical simulations we performed. We show this relative difference as a function of $\rho$, for different values of the non-reciprocity parameter $\delta$.}
\label{fig_difference_DA_DB}
\end{center}
\end{figure}

\begin{itemize}
    \item We show in Fig.~\ref{fig_Deff_A} the effective diffusion coefficient of $A$ particles. This is the counterpart of {Fig.~\ref{fig_Deff_B}}, 
    which showed the effective diffusion coefficient of $B$ particles. 

    \item We show in Fig.~\ref{fig_gr} the radial distribution functions for the different pairs of species in the binary mixtures (AA, BB, AB), and for the different interaction potentials used in the Brownian dynamics simulations (WCA-LJ, softcore, Yukawa). The strong pairing between A and B particles is visible on the functions $g_{AB}(r)$, which are much larger than $1$ when non-reciprocity is strong.

    \item We show in Fig.~\ref{fig_difference_DA_DB} the relative difference between the diffusion coefficients of $A$ and $B$ particles. More precisely, we define the diffusion coefficients rescaled by their reciprocal value: $\bar{D}^\alpha_\mathrm{eff}={D}^\alpha_\mathrm{eff}/D^\alpha_\mathrm{recip}$, and we define the relative difference between $\bar{D}^A_\mathrm{eff}$ and $\bar{D}^B_\mathrm{eff}$ as: $(\bar{D}_\mathrm{eff}^{B}-\bar{D}_\mathrm{eff}^{A})/\frac{1}{2}(\bar{D}_\mathrm{eff}^{A}+\bar{D}_\mathrm{eff}^{B})$. This quantity is plotted for the three sets of numerical simulations (LJ-WCA, softcore, Yukawa).

\end{itemize}

\section{Limit of soft interactions}
\label{app_lin_Dean}

In the main text, for simplicity, we considered the situation where all the particles have the same mobilities,  the same diffusion coefficients, and are connected to the same thermostat. We also assumed that the tracer was either a particle of species $A$, or a particle of species $B$. In this supplementary calculation, we consider the more general situation where the tracer, denoted by the index $0$, can be a different species than $A$ or $B$. There are \emph{a priori} three different thermostats ($T_0$, $T_A$ and $T_B$), three different mobilities ($\mu_0$, $\mu_A$ and $\mu_B$). The bare diffusion coefficients are related to temperatures and mobilities through the fluctuation-dissipation relation, which is assumed to hold in the limit of infinite dilution: $D_\alpha = \kB T_\alpha \mu_\alpha$. To ease the notation, we will denote by ``$0$'' the species of the tracer (which is different from $A$ and $B$ in the most general situation). Finally, although the results presented in the main text applied to a three-dimensional systems, we provide here a derivation that holds in any spatial dimension $d$.

\subsection{Linearized Dean equation}

We start from {Eq.~\eqref{Dean_A}}. We linearize the density fields $\hat\rho_\alpha$ around the homogeneous value $\rho_\alpha$, by writing $\hat\rho_\alpha= \rho_\alpha + \sqrt{\rho_\alpha}\psi_\alpha$ and assuming $\psi_\alpha \ll \sqrt{\rho_\alpha} $. We find, at leading order in $\psi_i$:
\begin{align}
\label{}
&\partial_t \psi_A =    \sqrt{2D_A} \nabla \cdot \boldsymbol{\eta}_A +D_A \nabla^2 \psi_A \nonumber\\
&-\rho \mu_A [  \nabla \cdot (X  {\FF_{A\to A}}\ast \psi_A) \nonumber\\
&+\nabla\cdot (\sqrt{X(1-X)}  {\FF_{B\to A}}\ast \psi_B) \nonumber\\
 &+ \frac{\sqrt{\rho_A}}{{\rho}} \nabla \cdot (  {\FF_{0\to A}}\ast \delta_{\rr_0})   ] ,
\end{align}
and the counterpart for $\psi_B$. For simplicity, we denoted $X_A=X$ and $X_B=1-X$.
In Fourier space, the coupled equations for $\tilde{\psi}_A(\qq,t)$ and $\tilde{\psi}_B(\qq,t)$ then read
\begin{align}
\label{sol_phi_Fourier}
& \partial_t 
 \begin{pmatrix}
     \tilde{\psi}_A(\qq,t)     \\
     \tilde{\psi}_B(\qq,t)   
\end{pmatrix}   = -\boldsymbol{m} \begin{pmatrix}
     \tilde{\psi}_A(\qq,t)     \\
     \tilde{\psi}_B(\qq,t)   
\end{pmatrix} 
+
\begin{pmatrix}
     \sqrt{2D_A}\ii \boldsymbol{q}\cdot \tilde{\boldsymbol{\eta}}_A     \\
      \sqrt{2D_B} \ii \boldsymbol{q}\cdot  \tilde{\boldsymbol{\eta}}_B
\end{pmatrix} \nonumber\\
 &-\rho
    \begin{pmatrix}
      \sqrt{\frac{X}{{\rho}}} {\ex{-\ii \qq \cdot \rr_0(t)}}\mu_A\ii \qq\cdot \tilde\FF_{0\to A}    \\
         \sqrt{\frac{1-X}{{\rho}}} {\ex{-\ii \qq \cdot \rr_0(t)}}\mu_B \ii \qq\cdot \tilde\FF_{0\to B} 
\end{pmatrix}  
,
\end{align}
with
\begin{multline}
\boldsymbol{m} =
\left(\begin{matrix}
  q^2 {\kB T_A}\mu_A+ \rho \mu_A X \ii \qq\cdot\tilde \FF_{A\to A} \\
    \rho \mu_B \sqrt{X(1-X)} \ii \qq\cdot \tilde\FF_{A\to B}\end{matrix}\right.
    \\
    \left.
  \begin{matrix}  
   \rho  \mu_A\sqrt{X(1-X)} \ii \qq\cdot\tilde \FF_{B\to A}  \\
   q^2 {\kB T_B}\mu_B +  \rho \mu_B(1-X) \ii \qq\cdot \tilde\FF_{B\to B}
\end{matrix}\right)
\label{m_explicite}
\end{multline}
We apply the same procedure to the equation for the position of the tracer, which reads
\begin{align}
\label{eq_tracer_lin}
& \frac{\dd}{\dd t}\rr_0(t) = \rho \mu_0 \sqrt{\frac{X}{{\rho}}} (\FF_{A \to 0}\ast\psi_A)(\rr_0(t),t) \nonumber\\
&+\rho \mu_0 \sqrt{\frac{1-X}{{\rho}}} (\FF_{B\to 0}\ast\psi_B)(\rr_0(t),t) +\sqrt{2D_0}\boldsymbol{\xi}(t)
\end{align}

In order to treat the coupling between the dynamics of the fields and that of the position of the tracer, we rely on a path-integral formulation, that was initially proposed in the situation where the tracer is coupled to a single fluctuating field \cite{Demery2011}, and later extended to the situation where the tracer is coupled to multiple fields \cite{Jardat2022}. In Section \ref{general_formalism}, we present a general formalism, that applies to a tracer particle coupled in a non-reciprocal way to two fluctuating fields, which themselves interact non-reciprocally.  In Section \ref{sec:mapping}, we show how the formalism can be applied to the case where the fluctuating fields represent the stochastic density fields of $A$ and $B$ particles.

\subsection{General formalism}
\label{general_formalism}

\begin{figure}[b]
\begin{center}
\includegraphics[width=0.8\columnwidth]{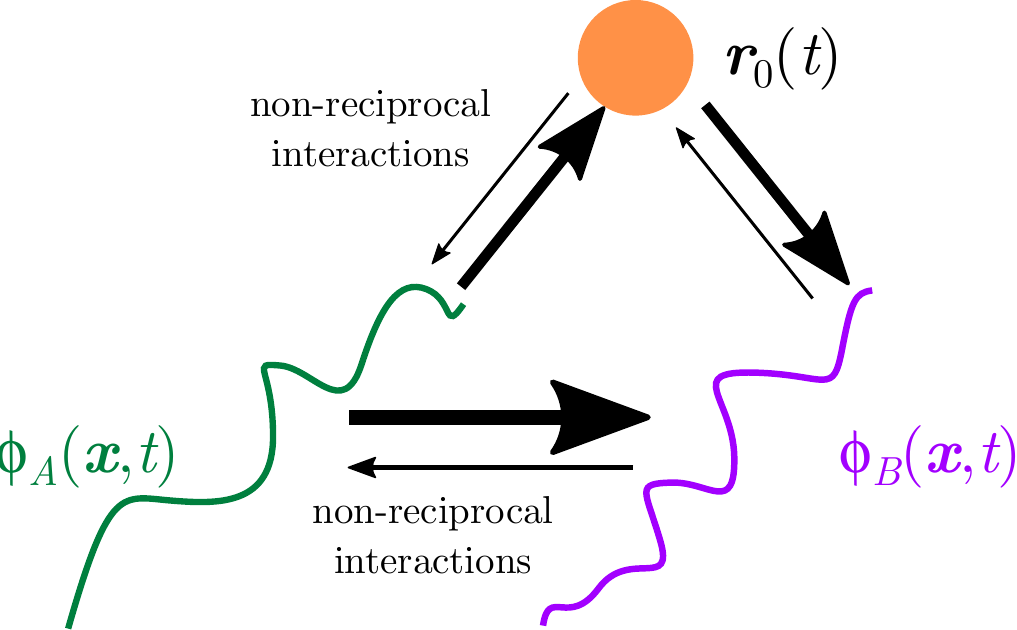}
\caption{ A diffusing tracer is coupled to two fluctuating fields, whose interactions may be non-reciprocal. The tracer-field interactions may also be non-reciprocal.}
\label{system}
\end{center}
\end{figure}

\subsubsection{Dynamics of the tracer and of the fields}

   We consider a tracer, whose position at time $t$ is denoted by $\rr_0(t)$, and which diffuses while being coupled to two fluctuating fields $\psi_A(\xx,t)$ and $\psi_B(\xx,t)$ (Fig.~\ref{system}). We assume that the interactions between the two fields can be non-reciprocal {and also that the tracer-field interactions can be non-reciprocal}. The position of tracer $\rr_0(t)$ obeys the following evolution equation:
  \begin{equation}
\frac{\dd }{\dd t} \rr_0(t) = \mu_0 \sum_{\alpha=A,B}\nabla { K_\alpha} \psi_\alpha[\rr_0(t)] +\sqrt{2D_0} \boldsymbol{\eta}(t) ,
\label{dyn_tracer}
\end{equation}
where $\mu_0$ is the mobility of the tracer, $h_\alpha$ is the coupling constant between the tracer and the field $\alpha$, and $K_\alpha$ is a linear operator.  Throughout the calculation we use the following shorthand notations for convolutions between operators and fields:
\begin{align}
\label{ }
AV(\xx) &= \int \dd \xx' \; A(\xx-\xx') V(\xx), \\
ABV(\xx) &= \int \dd \xx'\int \dd \xx''  \; A(\xx-\xx')B(\xx'-\xx'') V(\xx''). 
\end{align}
The noise term $ \boldsymbol{\eta}(t) $ has average zero and unit variance: $\moy{\eta_i(t)\eta_j(s)} =  \delta_{ij} \delta (t-s)$.

\emph{Dynamics of the fields.---} We then assume that the fields $\psi_A$ and $\psi_B$ obey the following dynamics:
 \begin{align}
&\partial_t \psi_A(\xx,t)  =  - R_A \left[ \Delta_{AA} \psi_A +   \Delta_{AB}\psi_{B}\right] \nonumber\\
&+ R_A  {K'_A}[\xx-\rr_0(t)] + \sqrt{2D_A}\xi_A(\xx,t) \label{phi_A_real} \\
&\partial_t \psi_B(\rr,t)  =  - R_B \left[   \Delta_{BA} \psi_A +  \Delta_{BB}\psi_{B}\right] \nonumber\\
&+ R_B  {K'_B}[\xx-\rr_0(t)] + \sqrt{2D_B}\xi_B(\xx,t) \label{phi_B_real}
\end{align}
where $\mu_\alpha$ is the mobility of the field $\alpha$, $R_\alpha$ is an operator used to specify if the dynamics is conservative or not (in Fourier space, $\tilde R_\alpha (\qq)=1$ corresponds to a non-conserved `model A' dynamics, whereas $\tilde R_\alpha(\qq)=\qq^2$ corresponds to a conserved `model B' dynamics \cite{Chaikin}).  { Note that, for non-reciprocal interactions, $K_\alpha \neq K'_\alpha$.} The noise terms are such that $\moy{\xi_\alpha(\xx,t)\xi_\beta(\xx',s)} =  \delta_{\alpha\beta} R_\alpha(\xx-\xx')\delta(t-s)$.

The first term in the rhs of Eqs. \eqref{phi_A_real} and \eqref{phi_B_real}, describes the interactions between the fields, whereas the second term describes the effect of the probe on the evolution of the fields. Note that, as opposed to the calculation proposed in \cite{Demery2011}, we do not assume that the dynamics of the system can be written as deriving from a Hamiltonian $\mathcal{H}[\rr_0(t),\psi_A,\psi_B]$. Indeed, this choice of dynamics  would yield reciprocal interactions between fields $A$ and $B$, by construction.

We then follow the lines of the calculation presented in Ref.~\cite{Jardat2022}, and we recall all the steps for completeness. The main difference is that the tracer-bath coupling are represented differently in Eq.~\eqref{dyn_tracer} and in Eqs. \eqref{phi_A_real}-\eqref{phi_B_real}: the reciprocal case would be recovered in the particular case $K'_A=K_A$ and $K'_B=K_B$.

\subsubsection{Generalized Langevin equation for the tracer}

 The next step of the calculation consists in deriving a generalized Langevin equation obeyed by the position of the tracer. To this end, we first solve for the dynamics of the fields $\psi_\alpha(\rr,t)$. We start from Eq.~\eqref{phi_A_real} and  \eqref{phi_B_real}. The equations for $\psi_A$ and $\psi_B$ read, in Fourier space:
\begin{align}
\label{Fourier1}
&\frac{\dd}{\dd t} \begin{pmatrix}
      \widetilde{\psi}_A(\qq,t)    \\
         \widetilde{\psi}_B(\qq,t) 
\end{pmatrix}
=
-\boldsymbol{m}
\begin{pmatrix}
      \widetilde{\psi}_A   \\
         \widetilde{\psi}_B
\end{pmatrix} \nonumber\\
&+
\begin{pmatrix}
    \ex{-\ii \qq \cdot \rr_0(t)}  \widetilde{R}_A  \widetilde{K}'_A  + \sqrt{2D_A} \widetilde{\xi}_A \\
   \ex{-\ii \qq \cdot \rr_0(t)}  \widetilde{R}_B  \widetilde{K}'_B  + \sqrt{2D_B} \widetilde{\xi}_B 
\end{pmatrix},
\end{align}
where the dependences over $\qq$ are not written explicitly for clarity, and where we define the matrix $\boldsymbol{m}$ as
\begin{equation}
\boldsymbol{m} = \begin{pmatrix}
   \widetilde{R}_A  \widetilde{\Delta}_{AA} &   \widetilde{R}_A\widetilde{\Delta}_{BA} \\
    \widetilde{R}_B  \widetilde{\Delta}_{AB}   &   \widetilde{R}_B  \widetilde{\Delta}_{BB}
\end{pmatrix}.
\label{expression_m}
\end{equation}
Eq.~\eqref{Fourier1} is a simple set of coupled linear first order differential equations, whose resolution requires the matrix exponential  $\widetilde{\boldsymbol{\mathcal{M}} } \equiv  \exp[-(t-s){\boldsymbol{m}}] $, which is written under the form
\begin{equation}
\label{matrix_exp}
\mathcal{M}_{\alpha\beta} = c^{(+)}_{\alpha\beta} \ex{-(t-s)\lambda_+}+c^{(-)}_{\alpha\beta}  \ex{-(t-s)\lambda_-},
\end{equation}
where we defined the matrices
\begin{equation}
\label{c_matrices}
\boldsymbol{c^{(\pm)} } =
\frac{1}{2s}
\begin{pmatrix}
{\pm m_{AA}\mp m_{BB}+s}
 &\pm 2{m_{AB}} \\
\pm 2{m_{BA}}
   & \mp m_{AA}\pm m_{BB}+s
\end{pmatrix},
\end{equation}
the eigenvalues
\begin{equation}
\lambda_\pm =  \frac{m_{AA}+m_{BB}}{2} \pm \frac{1}{2}\sqrt{(m_{AA}-m_{BB})^2+4m_{AB}m_{BA}},
\end{equation}
and the quantity
\begin{equation}
s = \sqrt{(m_{AA}-m_{BB})^2+4m_{AB}m_{BA}}.
\end{equation}
After Fourier inversion, one finds the solution of Eq.~\eqref{phi_A_real} in real space under the form 
\begin{align}
\label{sol_phi_alpha_real}
\psi_\alpha(\xx,t) =& \int_{-\infty}^t \dd s \sum_\beta \left\{ \mathcal{M}_{\alpha\beta}(t-s)R_\beta K'_\beta [\xx-\rr_0(s)]  \right. \nonumber\\
&\left.+ \sqrt{2D_{\beta}} \mathcal{M}_{\alpha\beta}(t-s) \xi_\alpha(\xx,s) \right\},
\end{align}
where $\mathcal{M}_{\alpha\beta}$ are the elements of the inverse Fourier transform of $\widetilde{\boldsymbol{\mathcal{M}} }$.

Starting from Eq.~\eqref{dyn_tracer} and using the expression for the field derived previously [Eq.~\eqref{sol_phi_alpha_real}], the equation for the dynamics of the tracer can be rewritten as
\begin{align}
\label{dyntracer2}
&\frac{\dd}{\dd t} \rr_0(t) = \sqrt{2D_0} \boldsymbol{\eta}(t) \nonumber\\
&+ \int_{-\infty}^t \dd s \;  \boldsymbol{F}[\rr_0(t)-\rr_0(s),t-s]+ \boldsymbol{\Xi}[\xx_0(t),t],
\end{align}
with
\begin{equation}
\label{def_F}
 \boldsymbol{F}(\rr,t)=\mu_0\sum_{\alpha,\beta}  \nabla K_\alpha \mathcal{M}_{\alpha\beta} (t) R_\beta K'_\beta (\rr),
\end{equation}
and
\begin{equation}
\label{ }
\boldsymbol{\Xi}[\rr,t] =\mu_0 \sum_{\alpha,\beta} \sqrt{2D_\beta} \nabla K_\alpha \int_{-\infty}^t \dd s\; \mathcal{M}_{\alpha\beta}(t-s) \xi_\beta(\rr,s)
\end{equation}
Therefore, the dynamics of the tracer [Eq.~\eqref{dyntracer2}] is formally written as  a generalized Langevin equation.

\subsubsection{Path-integral representation}

Starting from Eq.~\eqref{dyntracer2}, we now aim at calculating the mean-square displacement of the tracer at a given time $t_f$, defined as $\langle [\rr_0(t_f)-\rr_0(0)]^2 \rangle$, and the self-diffusion coefficient, defined as
\begin{equation}
\label{ }
D_\text{eff} = \lim_{t_f\to\infty} \frac{\langle [\rr_0(t_f)-\rr_0(0)]^2 \rangle}{2d t_f}
\end{equation}
 To this end, we follow the lines of Ref. \cite{Demery2011}, in which a perturbative path-integral study was outlined. Introducing a variable $\pp$ conjugated to the position of the tracer, the partition function associated to Eq.~\eqref{dyntracer2} can be written under the form $Z = \int \mathcal{D}\xx\,  \mathcal{D}\pp \; \ex{-S[\xx,\pp]} $, where the action $S[\xx,\pp]=S_0[\xx,\pp] + S_\text{int}[\xx,\pp]$ has the following contributions:
\begin{align}
&S_0[\xx,\pp] = - \ii \int \dd t \; p_i(t) \dot x_i(t) +D_0 \int \dd t \; p_i(t) p_i(t),\\
 &S_\text{int}[\xx,\pp]  =  \ii \int \dd t \, \dd s \;  p_i(t) F_i[\rr_0(t)-\rr_0(s),t-s] \theta (t-s) \nonumber\\
 &+ \int \dd t \, \dd s \; p_i(t) G_{ij}[\rr_0(t)-\rr_0(s),t-s] p_j(t)  \theta (t-s).
\end{align}
We used the Einstein summation convention and where $\theta$ denotes the Heaviside function. The matrix elements $G_{ij}$ are defined as $
G_{ij}(\xx-\xx',t-t') \equiv \langle \Xi_i(\xx,t) \Xi_j(\xx',t') \rangle$,
and read, in Fourier space:
\begin{align}
\label{ }
\tilde G_{ij}(\qq,t) = &2 \mu_0^2 q_i q_j \sum_{\alpha,\beta,\gamma} \tilde K_\alpha \tilde K_\gamma \kB T_\beta \tilde R_\beta  \nonumber\\
&\times\sum_{\nu,\epsilon=\pm1}  c_{\alpha\beta}^{(\nu)}c_{\gamma\beta}^{(\epsilon)} \frac{\ex{-\lambda_\nu |t|}}{\lambda_\nu + \lambda_\epsilon},
\end{align}
where the sums over $\alpha$, $\beta$ and $\gamma$ run over all the constituents of the mixture, and where we use the expression of the matrix exponential $\mathcal{M}_{\alpha\beta}$ given in Eq.~\eqref{matrix_exp}. Similarly, the Fourier transform of the components of $\FF$, defined in Eq. \eqref{def_F}, read
\begin{equation}
    \tilde F_{i}(\qq,t) = \ii \mu_0 \qq \sum_{\alpha\beta}\tilde{K}_\alpha \tilde {\mathcal{M}}_{\alpha\beta}(t) \tilde R_\beta \tilde K'_\beta.
\end{equation}
At equilibrium, when $\tilde K_\alpha=K'_\alpha$, $T_\alpha=T$ and $\Delta_{\alpha\beta}=\Delta_{\beta\alpha}$, one can check that the functions $\FF$ and $\GG$ satisfy the relation $\nabla_i F_j(\xx,t)=\partial_t G_{ij}(\xx,t)$ for $t>0$ \cite{Basu2022Dynamics}.

Expanding in the limit where the tracer-bath interactions are small (i.e. when the interaction action $S_\text{int}$ is small compared to $S_0$) and at first nontrivial order, one gets the following expression for the mean-square displacement of the tracer:
\begin{equation}
\label{ }
\langle [\rr_0(t_f)-\rr_0(0)]^2\rangle \simeq \langle [\rr_0(t_f)-\rr_0(0)]^2\rangle_0 -I_F-I_G,
\end{equation}
where the average $\langle \dots \rangle_0$ is taken over the bare action $S_0$, and where we defined 
\begin{align}
I_F    =& \Bigg \langle \ii \rr_0(t_f)^2 \int \dd t\int \dd s \; \theta(t-s) p_i(t) \nonumber\\
&\times F_{i}[\rr_0(t)-\rr_0(s),t-s] \Bigg\rangle_0  \\
      \underset{t_f\to\infty}{\simeq}& 4 D_0 \int \frac{\dd^d \qq}{(2\pi)^d} \;q^2 \mu_0  \sum_{\alpha,\beta} \tilde{K}_\alpha(\qq) \tilde{K}'_\beta(\qq) \tilde{R}_\beta(\qq) \nonumber\\
      &\times \sum_{\nu=\pm1}  \frac{c^{(\nu)}_{\alpha\beta}}{(D_0q^2+\lambda_\nu)^2}t_f,
\end{align}
and
\begin{align}
I_G    =& \Bigg \langle   \rr_0(t_f)^2 \int \dd t\int \dd s \; \theta(t-s) p_i(t) \nonumber\\
&\times G_{ij}[\rr_0(t)-\rr_0(s),t-s] p_j(s)\Bigg\rangle_0  \\
      \underset{t_f\to\infty}{\simeq}& 4  \int \frac{\dd^d \qq}{(2\pi)^d} \;q^2 \mu_0^2  \sum_{\alpha,\beta,\gamma}  \tilde{K}_\alpha(\qq) \tilde{K}_\gamma(\qq) \tilde{R}_\beta(\qq) \kB T_\beta \nonumber\\
&\times\sum_{\nu,\epsilon=\pm 1} \frac{c_{\alpha,\beta}^{(\nu)}c_{\gamma,\beta}^{(\epsilon)}}{\lambda_\nu+\lambda_\epsilon} \cdot \frac{D_0 q^2-\lambda_\nu}{(D_0 q^2+\lambda_\nu)^2} t_f.
\end{align}

Then, using the definition of the effective diffusion coefficient and integrating over all Fourier modes, we write the effective diffusion coefficient under the form $D_\text{eff} = D_0 - \sum_{\alpha,\beta} \bar{D}_{\alpha\beta}$
with
\begin{align}
\label{Dalphabeta}
 &\bar{D}_{\alpha\beta} = \frac{\mu_0 }{d}  \int \frac{\dd^d \qq}{(2\pi)^d} 
q^2 \tilde{K}_\alpha(\qq) \tilde{R}_\beta(\qq) \nonumber \\
&\times  \sum_\gamma \sum_{\nu = \pm 1}  \frac{c^{(\nu)}_{\alpha\beta}}{(D_0 q^2 + \lambda_\nu)^2}   \Bigg[  2D_0 \delta_{\gamma\beta} \tilde{K}'_\gamma(\qq) \nonumber\\
&+D_0  \left( \frac{c^{(\nu)}_{\gamma\beta}}{\lambda_\nu }+\frac{2c^{(-\nu)}_{\gamma\beta}}{\lambda_+ + \lambda_-}  \right) (D_0 q^2 - \lambda_\nu) \tilde{K}_\gamma(\qq)    \Bigg]   .
\end{align}

\subsection{Correction to the mobility}\label{app_corr_mob}

Here we adapt the calculation above to the correction to the mobility, following Ref.~\cite{Demery2014}.
We apply a small external force $\ff$ and compute the correction to the average tracer position, $\langle \rr_0(t_f)-\rr_0(0) \rangle$.
The bare action is now given by
\begin{align}
S_0[\xx,\pp] = &- \ii \int \dd t \; p_i(t) \left[\dot r_{0,i}(t)-\mu_0\ff\right] \nonumber\\
&+D_0 \int \dd t \; p_i(t) p_i(t),
\end{align}
The average displacement of the tracer is
\begin{equation}
\langle \rr_0(t_f)-\rr_0(0) \rangle = \mu_0\ff t_f - I'_F - I'_G,
\end{equation}
where the following averages are computed with Ref.~\cite{Demery2014} (Eqs.~(72, 73)):
    \begin{align}
I'_F  =& \Bigg \langle \ii \rr_0(t_f) \int \dd t\int \dd s \; \theta(t-s) p_i(t) \nonumber\\
&\times F_{i}[\rr_0(t)-\rr_0(s),t-s] \Bigg\rangle_0 \\
&= -\ii\mu_0 t_f\int \frac{\dd^d \qq}{(2\pi)^d} \sum_{\alpha,\beta}\sum_{\nu=\pm 1} \frac{\qq \tilde K_\alpha\tilde K'_\beta\tilde R_\beta c_{\alpha\beta}^{(\nu)}}{\lambda_\nu+D_0 q^2-\ii\mu_0\qq\cdot\ff}. 
\end{align}
and
\begin{align}
I'_G  = &\Bigg \langle   \rr_0(t_f) \int \dd t\int \dd s \; \theta(t-s) p_i(t) \nonumber\\
&\times G_{ij}[\rr_0(t)-\rr_0(s),t-s] p_j(s)\Bigg\rangle_0 \\
&= -2\ii\mu_0^2 t_f \int \frac{\dd^d \qq}{(2\pi)^d} \sum_{\alpha,\beta,\gamma}\sum_{\nu,\epsilon =\pm 1}\qq\, q^2\tilde K_\alpha\tilde K_\gamma \kB T_\beta\tilde R_\beta \nonumber\\
&\times \frac{c_{\alpha\beta}^{(\nu)}c_{\gamma\beta}^{(\epsilon)}}{(\lambda_\nu+\lambda_\epsilon)(\lambda_\nu+D_0 q^2-\ii\mu_0\qq\cdot\ff)}.
\end{align}

To obtain the effective mobility, we take the limit $f\to 0$ and consider the velocity in the direction of the force, leading to
\begin{align}
&\mu_\text{eff}=\mu_0
-\frac{\mu_0^2}{d} \int \frac{\dd^d \qq}{(2\pi)^d} \sum_{\alpha,\beta}\sum_{\nu=\pm 1} \frac{q^2 \tilde K_\alpha\tilde K'_\beta\tilde R_\beta c_{\alpha\beta}^{(\nu)}}{(\lambda_\nu+D_0 q^2)^2} \nonumber\\
&-\frac{2\mu_0^3 }{d}\int \frac{\dd^d \qq}{(2\pi)^d} \sum_{\alpha,\beta,\gamma}\sum_{\nu,\epsilon =\pm 1}\frac{q^4\tilde K_\alpha\tilde K_\gamma \kB T_\beta\tilde R_\beta c_{\alpha\beta}^{(\nu)}c_{\gamma\beta}^{(\epsilon)}}{(\lambda_\nu+\lambda_\epsilon)(\lambda_\nu+D_0 q^2)^2}\\
&=\mu_0
-\frac{\mu_0^2}{d} \int \frac{\dd^d \qq}{(2\pi)^d} \sum_{\alpha,\beta}\sum_{\nu=\pm 1} \frac{q^2 \tilde K_\alpha\tilde R_\beta c_{\alpha\beta}^{(\nu)}}{(\lambda_\nu+D_0 q^2)^2} \nonumber\\
&\times \left[\tilde K'_\beta + 2\mu_0   \kB T_\beta q^2 \sum_{\gamma}\sum_{\epsilon =\pm 1} \frac{\tilde K_\gamma  c_{\gamma\beta}^{(\epsilon)}}{\lambda_\nu+\lambda_\epsilon} \right].
\label{mobility_general}
\end{align}

\subsection{Mapping with the situation considered in the main text}
\label{sec:mapping}

The equations of the main text, which were derived in the specific situations where the fields $\psi_A$ and $\psi_B$ describe the perturbation around  homogeneous densities of  of interacting Brownian particles, are mapped onto the general equations [Eqs. \eqref{dyn_tracer} and \eqref{Fourier1}] using the following expressions for the operators $R_\alpha$, $\Delta_{\alpha\beta}$, $K_\alpha$ and $K'_\alpha$:
\begin{align}
&\tilde{R}_\alpha(\qq) =\mu_\alpha q^2 \label{mapping1}  \\
&\begin{cases}
 \tilde{\Delta}_{AA} = \kB T_A q^2 + \rho X \ii \qq\cdot\tilde{\FF}_{A\to A} \\
 \tilde{\Delta}_{BB} = \kB T_B q^2+ \rho(1-X)\ii \qq\cdot\tilde{\FF}_{B\to B}  \\
  \tilde{\Delta}_{AB}   = \rho\sqrt{X (1-X)}\ii \qq\cdot\tilde{\FF}_{B\to A}  \\
    \tilde{\Delta}_{BA}   = \rho\sqrt{X (1-X)}\ii \qq\cdot\tilde{\FF}_{A\to B}  \\
\end{cases} \label{mapping2}\\
&\begin{cases}
   \tilde{K}_A = -\frac{\sqrt{\rho X}}{q^2} \ii \qq\cdot\tilde{\FF}_{A\to 0} \\
   \tilde{K}_B = -\frac{\sqrt{\rho (1-X)}}{q^2} \ii \qq\cdot\tilde{\FF}_{B\to 0} 
\end{cases}\\
 &\neq  \begin{cases}
   \tilde{K}'_A =-\frac{\sqrt{\rho X}}{q^2} \ii \qq\cdot\tilde{\FF}_{0\to A} \\
   \tilde{K}'_B =-\frac{\sqrt{\rho (1-X)}}{q^2} \ii \qq\cdot\tilde{\FF}_{0\to B} 
\end{cases}.
\label{mapping4}
\end{align}

\subsection{Expression of the long-time diffusion coefficient and mobility}

Using the mapping from Eqs. \eqref{mapping1}-\eqref{mapping4}, we get the following expression of the effective diffusion coefficient of the tracer particle:
\begin{align}
&\frac{ {D}_\text{eff} }{D_0}=1-\sum_{\alpha,\beta,\gamma} {\mu_0 \mu_\beta}  \int_0^\infty \frac{\dd q}{6\pi^2}  q^2\rho\sqrt{X_\alpha X_\gamma}\ii \qq \cdot \tilde\FF_{\alpha \to 0} \nonumber\\
&\times \sum_{\nu=\pm1}   \frac{2 c^{(\nu)}_{\alpha\beta}}{(D_0 q^2 + \lambda_\nu)^2}   \Bigg[   \delta_{\gamma\beta} \ii \qq\cdot \tilde \FF_{0\to\gamma} \nonumber\\
&+ \frac{T_\beta}{T_0}(D_0 q^2 - \lambda_\nu) \sum_{\epsilon=\pm1}  \frac{c^{(\epsilon)}_{\gamma\beta}}{\lambda_\nu + \lambda_\epsilon}  \ii \qq\cdot \tilde\FF_{\gamma\to 0}  \Bigg]     ,
\end{align}
where $\lambda_{\pm 1}$ denote the eigenvalues of $\boldsymbol{m}$ , and where the coefficients $c^{(\pm 1)}_{\alpha\beta}$ are the elements of the matrices
\begin{equation}
\label{c_matrices_main}
\boldsymbol{c^{(\pm)} } =
\frac{1}{2s}
\begin{pmatrix}
{\pm m_{AA}\mp m_{BB}+s}
 &\pm 2{m_{AB}} \\
\pm 2{m_{BA}}
   & \mp m_{AA}\pm m_{BB}+s
\end{pmatrix},
\end{equation}
with \begin{align}
\label{ }
&s \equiv  \{[(q^2 {\kB T_A}\mu_A+ \rho \mu_A X \ii \qq\cdot \FF_{A\to A}) \nonumber\\
&- (q^2 {\kB T_B}\mu_B +  \rho \mu_B(1-X) \ii \qq\cdot \FF_{B\to B} )    ]^2   \nonumber\\
& + 4 \rho^2 \mu_A \mu_B X(1-X)(\ii \qq\cdot \FF_{A\to B})(\ii \qq\cdot \FF_{B\to A})\}^{1/2}.
\end{align}

The expression given of $D_\text{eff}/D_0$ in the main text {[Eq.~\eqref{main_result_Deff}]} is obtained by considering the simple case where all the particles have the same mobilities, the same diffusion coefficients, and are connected to the same thermostat. We also consider the particular case where the forces derive from pseudo-potentials, i.e. in Fourier space:
\begin{equation}
\FF_{\alpha\to\beta}(\qq)=-\ii \qq \tilde \phi_{\alpha\to\beta}(\qq) .
\end{equation}
This yields the expression from the main text:
\begin{align}
&\frac{ {D}_\text{eff} }{D_0} = 1-\sum_{\substack{\alpha,\beta,\gamma \\ \in \{A,B\}}} \int \frac{\dd q}{6\pi^2} \rho q^2  \tilde \phi _{\alpha\to 0}(q)  \nonumber\\
&\times \sum_{\gamma} \sqrt{X_\alpha X_\gamma} \left[ C_{0\to\gamma}^{\alpha\beta\gamma} \tilde \phi _{\alpha\to 0}(q)  +C_{\gamma \to 0}^{\alpha\beta\gamma} \tilde \phi _{\gamma \to 0}(q)   \right],
\end{align}
with
\begin{align}
C_{0\to\gamma}^{\alpha\beta\gamma} & = \delta_{\gamma\beta} \sum_{\nu = \pm 1} \frac{2 c_{\alpha\beta}^{(\nu)}}{(1+\bar\lambda_\nu)^2}, \\
C_{\gamma\to 0}^{\alpha\beta\gamma} & = \sum_{\nu,\epsilon = \pm 1} \frac{2 c_{\alpha\beta}^{(\nu)}c_{\gamma\beta}^{(\epsilon)}}{(1+\bar\lambda_\nu)^2(\bar\lambda_\nu+\bar\lambda_\epsilon)} (1-\bar\lambda_\nu),
\end{align}
where we defined $\bar \lambda_\nu = \lambda_\nu/(D_0 q^2)$.

Similary, using the mapping and the expression of the mobility [Eq. \eqref{mobility_general}], we get
\begin{align}
&\frac{ \mu_\text{eff} }{D_0}=1-\sum_{\alpha,\beta,\gamma} \mu_0   \int_0^\infty \frac{\dd q}{6\pi^2}  q^2\rho\sqrt{X_\alpha X_\gamma}\ii \qq \cdot \tilde\FF_{\alpha \to 0} \nonumber\\
&\times \sum_{\nu=\pm1}   \frac{ c^{(\nu)}_{\alpha\beta}}{(D_0 q^2 + \lambda_\nu)^2}   \Bigg[   \delta_{\gamma\beta} \ii \qq\cdot \tilde \FF_{0\to\gamma} \nonumber\\
&+ 2\mu_0 \kB T_\beta q^2 \sum_{\epsilon=\pm1}  \frac{c^{(\epsilon)}_{\gamma\beta}}{\lambda_\nu + \lambda_\epsilon}  \ii \qq\cdot \tilde\FF_{\gamma\to 0}  \Bigg]     .
\label{mueff_soft_general}
\end{align}

\section{Effective mobility of a probe coupled non-reciprocally to a single bath}

In this section, we explain the derivation of the effective mobility of a probe coupled non-reciprocally to a single bath (see {Eq.~\eqref{mueff_oneB} in the main text}). We rely on the derivation presented in  \cite{Demery2019}, in which the mean displacement of a probe linearly coupled to a generic fluctuating field, and submitted to a driven harmonic confinement, is computed. In Eq.~(A.3) from this reference, the terms in brackets in the integrand has two contributions: the first one originates from memory effects, and should be proportional to the effect of the bath on the tracer times that of the tracer on the bath; the second one originates from the noise of the bath, and should be proportional to the square of the effect of the bath on the tracer.

Denoting by $(1-\delta)$ the effect of the bath on the tracer, as in the main text, and taking the limit of a harmonic trap of vanishing stiffness and vanishing velocity, we  find from Eq.~(A.3) in \cite{Demery2019}:
\begin{equation}
\mu_\text{eff} = \mu_0 -\frac{1}{3}\int\frac{\dd\qq}{(2\pi)^3} \frac{q^2\tilde K^2}{\tilde A} \frac{(1-\delta)((1-\delta)D_0 q^2+\tilde R\tilde A)}{(D_0 q^2+\tilde R\tilde A)^2},
\end{equation}
Using the following mapping between the notations of \cite{Demery2019} and our notations: $\tilde K^2 = \rho \tilde v^2$, $\tilde A = 1+\rho \tilde v$, $\tilde R  = \mu_0 q^2$, we find the result given in the main text. {This expression can also be obtained as a consequence of the more general expression of the long-time mobility [Eq. \eqref{mobility_general}].}

\section{Low-density limit}
\label{app_low_dens}

\subsection{Friction}\label{}

In order to measure the friction of the tracer particle in the low-density limit, we assume that it is submitted to a small external force, along the lines of \cite{Lekkerkerker1984}. Applying a force $\FF$ on the probe disturbs the pair distribution function $\gab(\rr)$, which is now a solution of
\begin{equation}
0 = 2D_0\nabla\cdot \left[ e^{-\uab}\nabla \left(e^{\uab}\gab \right) \right]+\mu_0\FF\cdot\nabla\gab,
\end{equation}
where we recall the definition of $\uab$ given in the main text: $\uab(r)= [\Phiab(r)+\Phiba(r)]/2\kB T$.
We use the ansatz
\begin{equation}\label{eq:g_fric}
\gab(\rr) = \gab^0(r) \left[1+q(r)\FF\cdot\hat\rr \right],
\end{equation}
where $\gab^0(r) $ is the pair distribution function in the absence of external force. In the limit of small force, the first order in $\boldsymbol{F}=F\hat{\boldsymbol{F}}$ reads
\begin{equation}
0 = 2D_0\nabla\cdot \left[ e^{-\uab(r)}\nabla \left(q(r)\hat\FF\cdot\hat\rr \right) \right]+\mu_0\hat \FF\cdot\nabla e^{-\uab(r)}.
\end{equation}
Writing explicitly the derivative and using $D_0=\kB T \mu_0$, we get
\begin{equation}\label{eq:def_q}
2\nabla\cdot \left[ e^{-\uab(r)}\nabla \left(q(r)\hat\FF\cdot\hat\rr \right) \right]
=\frac{1}{\kB T}\hat \FF\cdot\hat\rr \uab'(r) e^{-\uab(r)}.
\end{equation}
The friction created by the bath particles is
\begin{equation}
\FF\ind{b} = \sum_\beta \rho_\beta\int \gab(\rr)\nabla\Phiba(r)\dd\rr.
\end{equation}
Using the ansatz (\ref{eq:g_fric}) leads to
\begin{align}
\FF\ind{b} & = \sum_\beta\rho_\beta\int e^{-\uab(r)}q(r)(\FF\cdot\hat \rr)\hat\rr\Phiba'(r)\dd\rr\\
& = \sum_\beta\frac{\rho_\beta}{3}\FF \int e^{-\uab(r)}q(r)\Phiba'(r)\dd\rr
\end{align}
The relative change in mobility is thus
\begin{equation}\label{eq:correc_mob}
\frac{\mu\ind{eff}-\mu_0}{\mu_0} = \sum_\beta\frac{\rho_\beta}{3} \int e^{-\uab(r)}q(r)\Phiba'(r)\dd\rr.
\end{equation}

\subsection{Long-time diffusion coefficient}

From Eq.~(S21) in Ref.~\cite{Ilker2021}, the correction to the long time diffusion coefficient of the probe is given by
\begin{equation}\label{eq:correc_diff_vir_0}
\Delta D = -\lim_{s\to 0}\lim_{k\to 0} \left\langle e^{-i\qq\cdot\rr_1}\cL_N(s-\cL_N)^{-1}\cL_N e^{i\qq\cdot\rr_1} \right\rangle\ind{ss}.
\end{equation}
The average is defined by $\langle f \rangle\ind{ss}=\int f(\XX) P_N(\XX)\dd\XX$ where $\XX$ is the vector containing the coordinates of all the particles and $P_N(\XX)$ is the steady state probability density: $\cL_NP_N=0$. The Liouville operator is given by
\begin{align}
\cL_N f(\XX) =& D_0\sum_{n=1}^N\nabla_{\rr_n}\cdot \Bigg[\nabla_{\rr_n} f \nonumber\\
&+f\sum_{m\neq n} \nabla_{\rr_n}\phi_{S(m)\to S(n)}(\rr_n-\rr_m) \Bigg].
\end{align}
We now compute $\cL_N e^{i\qq\cdot\rr_1}P_N(\XX)$.
First, as $\cL_N P_N=0$, at least one derivative should act on $e^{i\qq\cdot\rr_1}$.
Second, as the limit $k\to 0$ will be taken, at most one derivative should  act on $e^{i\qq\cdot\rr_1}$.
We are left with
\begin{align}
&\cL_N e^{i\qq\cdot\rr_1}P_N(\XX) = D_0 e^{i\qq\cdot\rr_1} i\qq\cdot\Bigg[2 \nabla_{\rr_1} P_N(\XX)   \nonumber\\
&+ P_N(\XX) \nabla_{\rr_1} \sum_{m\neq 1}\phi_{S(m)\to S(1)}(\rr_1-\rr_m) \Bigg].
\end{align}
To simplify, we consider only two particles; we will multiply by the density afterwards.
We also take $P_N(\XX)=\gab^0(\rr_1-\rr_2)=\exp(-\uab(\rr_1-\rr_2))$, leading to
\begin{align}
&\cL_N e^{i\qq\cdot\rr_1}P_N(\XX)  \nonumber\\
&= -D_0e^{i\qq\cdot\rr_1- \uab(\rr_1-\rr_2)} i\qq\cdot[2 \nabla_{\rr_1} \uab(\rr_1-\rr_2)   \nonumber\\
&- \nabla_{\rr_1} \phiba(\rr_1-\rr_j) ]\\
&= -D_0e^{i\qq\cdot\rr_1- \uab(\rr)} i\qq\cdot\nabla_{\rr}\left[2\uab(\rr) - \phiba(\rr) \right]\\
&= -D_0e^{i\qq\cdot\rr_1- \uab(\rr)} i\qq\cdot\hat\rr\phiab'(r),
\end{align}
where $\rr=\rr_1-\rr_2$. Integrating by parts on the left in Eq.~(\ref{eq:correc_diff_vir_0}) and taking the limit $k\to 0$ leads to
\begin{equation}
\Delta D = -\sum_\beta\mu_0^2\rho_\beta I_{\alpha\beta}
\end{equation}
with
\begin{align}
I_{\alpha\beta}  =& \lim_{s\to 0} \int (\hat\qq\cdot\hat\rr)\Phiba'(r)(s-\cL_{\alpha\beta})^{-1} \nonumber\\
&\times \gab^0(r)\Phiab'(r) (\hat\qq\cdot\hat\rr)\dd\rr.
\end{align}
We introduce
\begin{equation}
\chiab(\rr,t=0) = \gab^0(r)\Phiab'(r) (\hat\qq\cdot\hat\rr),
\end{equation}
so that
\begin{equation}
\lim_{s\to 0} (s-\cL_{\alpha\beta})^{-1} \chiab(\rr,t=0) = \tchiab(\rr,s=0).
\end{equation}
The Laplace transform is solution of
\begin{align}
\cL_{\rr} \tchiab(\rr,s=0) &= -\chiab(\rr,t=0) \nonumber\\
&= -e^{-\uab(r)}\Phiab'(r) (\hat\qq\cdot\hat\rr).
\end{align}
Taking the ansatz
\begin{equation}\label{eq:ansatz_tchi}
\tchiab(\rr,s=0) = \gab^0(r)\Xab(r)\hat\qq\cdot\hat\rr,
\end{equation}
$\Xab(r)$ is the solution of
\begin{align}
&2\nabla\cdot \left[e^{-\uab(r)}\nabla_r\left(\Xab(r)\hat\qq\cdot\hat\rr\right) \right]  \nonumber\\
&= -D_0^{-1}e^{-\uab(r)}\Phiab'(r) (\hat\qq\cdot\hat\rr).
\end{align}
Plugging the ansatz (\ref{eq:ansatz_tchi}) in the correction, we get
\begin{equation}
I_{\alpha\beta} = \frac{1}{3}\int \Phiba'(r)e^{-\uab(r)}\Xab(r)\dd\rr,
\end{equation}
so that finally
\begin{equation}
\frac{D\ind{eff}-D_0}{D_0} = -\sum_\beta \frac{\mu_0\rho_\beta}{3\kB  T}\int \Phiba'(r)e^{-\uab(r)}\Xab(r)\dd\rr.
\end{equation}
Finally, defining $\Yab=-\mu_0\Xab/\kB T$, the correction and the definition of $\Yab(r)$ read
\begin{equation}\label{eq:correc_diff}
\frac{D\ind{eff}-D_0}{D_0} = \sum_\beta\frac{\rho_\beta}{3}\int \Phiba'(r)e^{-\uab(r)}\Yab(r)\dd\rr,
\end{equation}
and
\begin{equation}\label{eq:def_y}
2\nabla\cdot \left[e^{-\uab(r)}\nabla_r\left(\Yab(r)\hat\qq\cdot\hat\rr\right) \right] = \frac{1}{\kB T} \hat\qq\cdot\hat\rr \phiab'(r) e^{-\uab(r)}.
\end{equation}
We note that Eqs.~(\ref{eq:def_y}, \ref{eq:correc_diff}) are analogous to Eqs.~(\ref{eq:def_q}, \ref{eq:correc_mob}) for the effective mobility, the only change being that $\uab'(r)$ in the r.h.s of Eq.~(\ref{eq:def_q}) is replaced by $\phiab'(r)$ in Eq.~(\ref{eq:def_y}).
At equilibrium, $\phiba=\uab$ and the Einstein relation is recovered.

\subsection{Numerical evaluation of the correction}
\label{app_numerical_ode}

Equations (\ref{eq:def_q}) and (\ref{eq:def_y}) defining $q(r)$ and $\Yab(r)$ cannot be integrated analytically in general. 
We rewrite the equations here in a dimensionless form. 
The correction to the correlation satisfies
\begin{equation}
\nabla\cdot\left[e^{-u(r)}\nabla \left(q(r)\nn\cdot\hat\rr \right)\right] = \nn\cdot\hat\rr v'(r) e^{-u(r)},
\end{equation}
where $u(r)=\uab(r)$ and $v(r)=\uab(r)$ for the mobility and $v(r)=\phiab(r)$ for the diffusion coefficient; the vector $\nn$ is arbitrary.
{In arbitrary dimension $d$}, expanding this equation leads to
\begin{equation}
q''+ \left(\frac{d-1}{r}-u' \right)q'-\frac{d-1}{r^2}q = v'.
\end{equation}
The correction to the quantity $A$ ($A=\mu$ or $A=D$, depending on the function $v(r)$ used to compute $q(r)$) is then given by
\begin{equation}
\frac{\Delta A}{A} = \frac{S_{d-1}\bar\rho_\beta}{2d}\int_0^\infty e^{-u(r)}w'(r)q(r)r^{d-1}\dd r,
\end{equation}
where $w(r)=\phiba(r)$, {and $S_d$ denotes the surface of the $d$-dimensional unit sphere.}

Instead of solving for $q(r)$, we introduce $p=q e^{-u}$, which is the solution of
\begin{equation}\label{eq:ode_p}
p''+ \left(\frac{d-1}{r}+u' \right)p'+\left(u''+\frac{d-1}{r}u'-\frac{d-1}{r^2} \right)p = v'e^{-u}.
\end{equation}
With $p$, the correction reads
\begin{equation}
\frac{\Delta A}{A} = \frac{S_{d-1}\bar\rho_\beta}{2d}\int_0^\infty w'(r)p(r)r^{d-1}\dd r.
\end{equation}

For the correction to the pair correlation to be continuous at zero (Eq.~(\ref{eq:g_fric})), the solution should satisfy $p(0)=0$.
If the potentials diverge strongly at the origin, this condition can be replaced by $p(\epsilon)=0$, with small enough $\epsilon>0$.
If the potentials vanish beyond $r_c$, solving Eq.~(\ref{eq:ode_p}) for $u=v=0$ leads to $p(r)\propto r^{1-d}$ for $r>r_c$. 
This relation can be turned into the relation $p'(r)=\frac{1-d}{r}p(r)$, which may be used as the second boundary condition at $r=r_c$.
In the case where the potentials do not vanish beyond a given distance, such as LJ potentials, this boundary condition can still be used with a large enough $r_c$.
These two boundary conditions with the ODE (\ref{eq:ode_p}) form a boundary value problem, which may be solved numerically.


%

\end{document}